\documentclass[12pt]{article}
\usepackage{epsfig,multicol}
\usepackage[english]{babel}
\usepackage{graphicx}
\usepackage{amsmath}
\usepackage{amssymb}
\usepackage{flafter}
\usepackage{picinpar}

\newcommand{\code}[1]{{\tt #1}}

\newcommand{\ds}{{\sffamily DarkSUSY}}

\def\msun{M_{\odot}{\ }}

\begin{document}

\begin{titlepage}

\begin{flushright}
SISSA 95/2004/EP\\
FSU--HEP--041122
\end{flushright}

\begin{center}

\vspace*{2cm}
{\huge Neutralino Dark Matter Detection\\[0.5cm] in Split Supersymmetry Scenarios}\\
\vspace*{1.5cm}
{\bf\large A.~Masiero$^1$, S.~Profumo$^{2,3}$ and P.~Ullio$^3$}\\[0.7cm]
{\em $^1$ Dipartimento di Fisica `G.~Galilei', Universit\`a di
Padova, and Istituto Nazionale di Fisica Nucleare, Sezione di
Padova,
Via Marzolo 8, I-35131, Padova, Italy}\\[0.3cm]
 {\em $^2$ Department of Physics, Florida State University\\
505 Keen Building, FL 32306-4350, U.S.A.}\\[0.3cm]
 {\em $^3$ Scuola Internazionale Superiore di Studi Avanzati,
Via Beirut 2-4, I-34014 Trieste, Italy and Istituto Nazionale di
Fisica Nucleare, Sezione di Trieste,
I-34014 Trieste, Italy}\\[0.7cm]
{\em E-mail:} {\tt antonio.masiero@pd.infn.it, profumo@hep.fsu.edu,
ullio@sissa.it}

\vspace*{0.8cm}

\begin{abstract}
\noindent 
We study the phenomenology of neutralino dark matter
within generic supersymmetric scenarios where the Gaugino and
Higgsino masses are much lighter than the scalar soft breaking
masses (Split Supersymmetry). We consider a low-energy 
model-independent approach and show that the guidelines in the definition of 
this general framework come from cosmology, which forces the lightest neutralino 
to have a mass smaller than 2.2~TeV. The testability of the framework
is addressed by discussing all viable dark matter detection techniques.
Current data on cosmic rays antimatter, gamma-rays and on the abundance of primordial 
${}^6$Li already set significant constraints on the parameter space. 
Complementarity among future direct detection experiments, indirect searches 
for antimatter and with neutrino telescopes, and tests of the theory at future 
accelerators, such as the LHC and a NLC, is highlighted. In particular,
we study in detail the regimes of Wino-Higgsino mixing and Bino-Wino transition,
which have been most often neglected in the past. 
We emphasize that our analysis may apply to more general supersymmetric
models where scalar exchanges do not provide the dominant contribution to
annihilation rates.
\end{abstract}
\end{center}

\end{titlepage}
\section{Introduction}

Soon after the first explicit formulations of supersymmetric (SUSY) versions of the 
Standard Model (SM)~\cite{susyfirst}, enforcing R-parity conservation to prevent sources 
of violation of baryon and lepton number at tree-level, it was realized, 
as an interesting by-product, that these theories could naturally embed a 
candidate for cold dark matter~\cite{lspcdm}. In fact, it had been shown long 
before that weakly-interacting massive particles (WIMPs) have a thermal 
relic abundance that is generically at a level relevant for cosmology, and
the stable LSP falls under this category if it happens to be electric and color 
neutral (and as it is indeed the case in large portions of the SUSY parameter 
space).

Two decades later, no SUSY particle has unfortunately been found at 
particle accelerators and no clean indication of SUSY has emerged from tests
of rare processes or precision measurements. The initial theoretical motivations for 
SUSY, although still very appealing, are being questioned, with the scale of 
physics beyond the SM that is slowly, but steadily, drifting above the
weak scale. On the other hand, in these twenty years, the case for non-baryonic
cold dark matter (CDM) being the building block of the Universe has become 
stronger and stronger, with the determination of the CDM contribution to the Universe 
energy density which has already reached 
a 20\% level of accuracy~\cite{Spergel:2003cb} and is going to improve further in the 
upcoming years. It is then not surprising that in formulating an extension to the 
SM, sometimes, the issue of incorporating a dark matter candidate 
has changed from being a by-product of the proposed scenario into being one 
of the hinges of the theory itself. This is the case for a recently proposed
SUSY framework, dubbed "Split Supersymmetry"~\cite{Arkani-Hamed:2004fb,Giudice:2004tc}.

Split Supersymmetry indicates a generic realization of the SUSY extension to the
SM where fermionic superpartners feature a low mass spectrum 
(say at the TeV scale or lower), while scalar superpartners are heavy, with a mass 
scale which can in principle range from hundreds of TeV up to the GUT or the 
Planck scale \cite{Arkani-Hamed:2004fb}. In Split SUSY one gives up on the
idea that SUSY is the tool to stabilize the weak scale, since some other mechanism 
should anyway account for other fine tuning issues, such as that of the 
cosmological constant. On the other hand, a number of 
phenomenological problems appearing in ordinary SUSY setups, where the whole spectrum of superpartners is supposed to be light,
are cured: The heavy sfermions minimize flavor and CP violating effects, alleviate 
proton decay and avoid an excessively light Higgs boson. At the same time,
two major features are maintained: the successful
unification of gauge couplings, and -- the central issue in our analysis -- the LSP 
as a viable particle candidate for CDM. Finally, 
it is certainly a well-motivated scenario, as it has been shown that the occurrence 
of a Split SUSY spectrum is indeed a generic feature of a wide class of theories \cite{Arkani-Hamed:2004yi,Antoniadis:2004dt,Kors:2004hz,greco}.
It arises, among others, in frameworks where the mechanism of SUSY breaking 
preserves a R-symmetry and forbids Gaugino and Higgsino mass terms, or
in case of direct SUSY breaking through renormalizable interactions with the 
SUSY breaking sector (direct mediation) \cite{Kumar:2004yi,Hewett:2004nw,Arkani-Hamed:2004yi}. 

Rather than focusing on some specific realizations, we consider here a generic 
Split SUSY setup: we refer to the minimal SUSY extension to the Standard Model
(MSSM), set the scalar sector at a heavy scale, but allow for the most general 
fermion sector. We investigate how cosmological constraints affect the
definition of such framework and discuss its testability by 
examining  the issue of dark matter detection within this wide class of models,
comparing also with the perspectives to test this scenario at 
accelerators. As a result, we work out here a fully general analysis of neutralino dark matter on the basis of the lightest neutralino mass and composition, largely independent of the Split Supersymmetry framework; in this respect, we outline a reference guide to the sensitivity of upcoming detection experiments which applies, especially as fas as indirect detection is concerned, to supersymmetric models with a generic (moderatly) heavy scalar sector. Once the LSP composition and mass is known, our results allow to get a fair estimate of the detection prospects.

The paper is organized as follows: In Section~\ref{sec:paramspace} we describe 
our working framework and discuss the current constraints to its parameter space.
In Section~\ref{sec:rates} we introduce the relevant dark matter detection 
techniques and show rates in extremal regimes of the parameter space. In 
Section~\ref{sec:casestudy} we discuss in detail the phenomenology 
of the Split SUSY setup along a sample slice in the parameter space, 
show constraints from current data and address detectability in future
searches, cross-checking the outreach of the different techniques.
In Section~\ref{sec:discussion} we indicate how to generalize the results of 
the previous Section to the full parameter space. Finally, we draw an outlook and conclude in Section~\ref{sec:conclusion}.

\section{Definition of the model and parameter space}
\label{sec:paramspace}

We consider a MSSM setup, with the relevant parameters defined at
at the low energy scale and without implementing any 
unification scheme. We restrain to the Split SUSY framework by assuming
that the sfermion sector is much heavier than the fermion sector and hence 
is completely decoupled; as for the Higgs sector, we assume there is only one light SM-like Higgs, with a mass labeled by the parameter $m_h$,
while the other two neutral Higgs and the charged Higgs are taken again 
 to be very heavy, and hence decoupled. Finally, we will leave out of most of our discussion the 
Gluino and the Gravitino, supposing they are (moderately) heavy, just
mentioning what are possible consequences in case this is not true. 

In this scheme, the lightest neutralino, 
defined as the lightest mass eigenstate from the superposition of the two 
neutral Gaugino and the two neutral Higgsino fields,
\begin{equation}
  \tilde{\chi}^0_1 = 
  N_{11} \tilde{B} + N_{12} \tilde{W}^3 + 
  N_{13} \tilde{H}^0_1 + N_{14} \tilde{H}^0_2\;,
\end{equation}
is automatically the LSP\footnote{The lightest chargino is always heavier than the lightest neutralino at 
tree level; this mass hierarchy is moreover stable, in the parameter space
regions under scrutiny, against the inclusion of radiative corrections in 
the relevant mass matrices~\cite{radiativecorr}.}.
The coefficients $N_{1j}$, obtained by diagonalizing the neutralino mass matrix, 
are mainly a function of the Bino and the Wino mass parameters $M_1$ and 
$M_2$, and of  the Higgs superfield parameter $\mu$, while depending rather
weakly on $\tan\beta$, the ratio of the vacuum expectation values of the two 
neutral components of the SU(2) Higgs doublets. The hierarchy between
$M_2$ and $\mu$ determines also whether the lightest chargino is Wino like
or Higgsino like, and again the role $\tan\beta$ is minor. 
The phenomenology of the scheme we are considering is hence fully defined 
by only five parameters:
\begin{equation}
 M_1,\;M_2,\;\mu,\;\tan\beta\;\;\;{\rm and}\;\;\;m_h.
\end{equation}

In this construction, constraints from accelerator data are rather weak, 
essentially just in connection with the fact that charginos have not been 
observed so far: we implement, as a conservative limit on the lightest 
chargino mass, the kinematic limit from the last phase of LEP2, {\em i.e.} 
$m_{\widetilde\chi^+}>$103.5~GeV \cite{lep2} (notice however that in large portions of our parameter space the lightest 
neutralino and lightest chargino are nearly degenerate in mass, and this
limit should be slightly relaxed; however, this does not critically enter in 
our discussion). 

\begin{figure*}[!t]
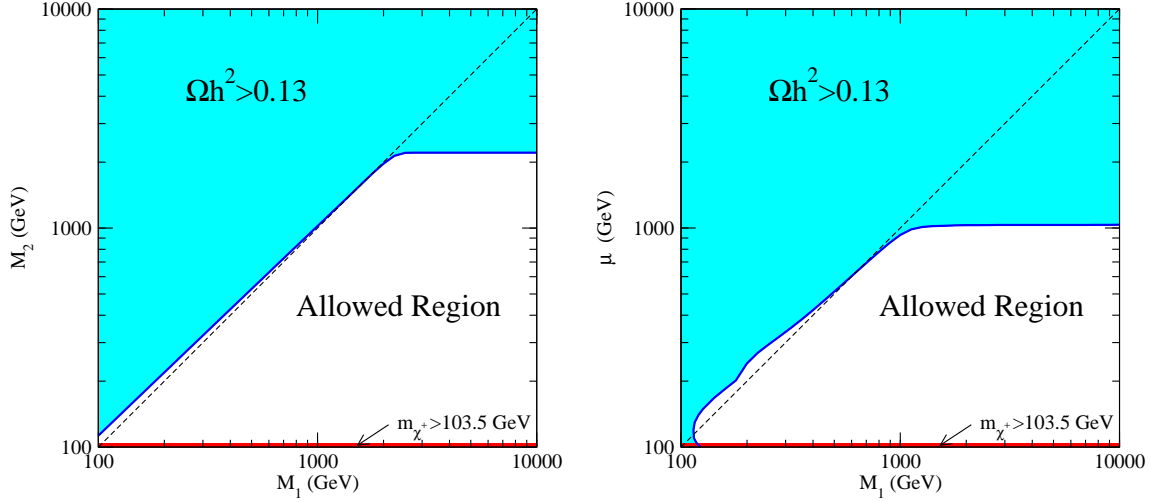

\begin{center}
\hspace*{-0.5cm}\includegraphics[scale=0.45]{plots/m1m2.eps}\quad\includegraphics[scale=0.45]{plots/m1mu.eps}\\
\end{center}
\caption{\small \em The parameter space region in the $(M_1,M_2)$
and $(M_1,\mu)$ plane allowed by direct chargino searches at LEP
and by the requirement that the neutralino relic abundance does
not exceed the CDM density. The third parameter (respectively $\mu$ (left) and $M_2$ (right)) is always set to 10~TeV. The excluded regions are respectively
shaded in red and light blue.}\label{m1m2}
\end{figure*}
\begin{figure*}[!h]
\begin{center}
\hspace*{-0.5cm}\includegraphics[scale=0.45]{plots/m2mu.eps}\qquad\qquad\includegraphics[scale=0.62]{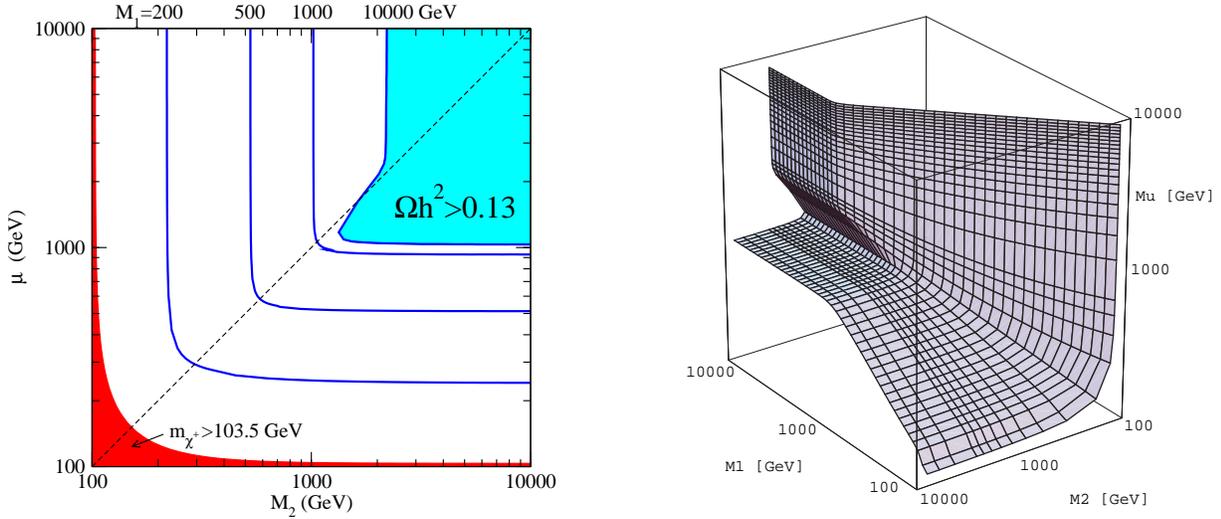}\\
\end{center}
\caption{\small \em (left): The allowed regions in the
$(M_2,\mu)$, with the same color code as in Fig.~\ref{m1m2}; the
cosmological limit is indicated for various values of $M_1$.
(right): A 3-D representation of the hypersurface at $\Omega_\chi
h^2 =0.113$ in the $(M_1,M_2,\mu)$ space.}\label{3Dlabels}
\end{figure*}

Much more powerful are constraints on the model from 
cosmology; to implement the cosmological bounds, 
we refer to the determination of  the CDM component of the Universe by  WMAP~\cite{Spergel:2003cb}: $\Omega_{CDM} h^2 = 0.113 \pm 0.009$. 
LSP relic densities are computed with the \ds\ package~\cite{Gondolo:2004sc}, which 
allows for high accuracy solutions of the equations describing thermal 
decoupling. We study the parameter space varying freely the mass parameters 
$M_1$, $M_2$ and $\mu$ (no GUT relations are assumed between $M_1$ 
and $M_2$; we restrict to positive $\mu$, since in this contest the case of 
negative $\mu$ is just specular), and taking the sample value $\tan\beta=50$ and 
$m_h = 115$~GeV (close to the current lower limit on the mass of a SM-like 
Higgs).

In Fig.~\ref{m1m2} we consider slices, in the parameter space, along
the planes $(M_1,M_2)$ (left-hand-side panel) and $(M_1,\mu)$ 
(right-hand-side panel) and for large values of the third mass parameter
(which we set to 10~TeV). In the first case, since $M_1$ and $M_2$ are both diagonal
entries in the neutralino mass matrix, the transition of the LSP from 
being Bino like (above the diagonal in the Figure) to being Wino like
(below it) is very sharp. The two regimes are very different because Winos
annihilate very efficiently into gauge bosons, while Binos can annihilate 
just into fermions. The latter annihilation processes are very sharply suppressed, as compared 
to ordinary SUSY setups, since diagrams with $t$- and $u$-channel sfermions
exchanges do not contribute here. 
In general, the relic density of the LSP scales with
the inverse of the annihilation rate: it is very large for Binos regardless
of the LSP mass, while it tends to be too small for Winos, unless one 
considers  $M_2$ in the 2~TeV range. The exception is in a narrow strip 
above the diagonal where the Bino LSP is nearly degenerate in mass with
the next-to-lightest Wino like neutralino and chargino states, and the 
thermal equilibrium of the LSP in the early Universe is enforced through
these slightly heavier particles rather than by just LSP pair annihilations
into SM particles: this is a very well-studied effect, usually dubbed
"coannihilation" ~\cite{coann}, but that has been very seldom considered 
in this specific realization, since usually $M_1$ and $M_2$ are not taken 
as independent parameters. The region in the plane 
where the LSP relic abundance is too large is shaded. Starting from
light LSPs (the mass of the LSP is essentially the minimum between 
$M_1$ and $M_2$) in the bottom-left corner, the border of the excluded
region approaches the diagonal, since increasing the mass scale the 
cross sections decreases, and hence coannihilation effects have to become 
larger and larger, 
{\em i.e.} mass splitting with coannihilating particles smaller and smaller.
We reach then the mass value at which this process is saturated, and the 
LSP turns into Wino like; above this scale, even for Winos, the annihilation
cross section becomes too small and CDM is overproduced. 
In the Figure, since we are plotting results on logarithmic scales,
the iso-level curve at  $\Omega_{CDM} h^2 = 0.113$ is essentially 
overlapping with the border of the excluded region, while in the remaining
part of the plane, CDM is underproduced. In Fig.~\ref{m1m2} and \ref{3Dlabels} 
we shade in red the regions with an excessively light chargino.

For the right-hand-side panel of Fig.~\ref{m1m2},  the picture is analogous.
The Wino mass parameter is assumed to be heavy, hence we are left with
the possibility that the LSP is either Bino like or Higgsino like, depending on 
the relative values of $M_1$ and $\mu$. Being the parameter $\mu$
in off-diagonal entries in the neutralino mass matrix, the transition between
the two regimes is smoother than in the previous case, but
the trend is the same, with Higgsinos annihilating efficiently
into gauge bosons and Binos with suppressed annihilation rates.
In the computation
of the relic density both the effects of mixing and coannihilations enter, and,
for instance, a wiggle in the borderline of the excluded region 
corresponding to the $t-\bar{t}$ threshold is clearly visible. This is the scenario which has been
considered up to now for estimates of the LSP relic abundance in Split SUSY 
contexts~\cite{Giudice:2004tc,Pierce:2004mk,Arkani-Hamed:2004yi}.

The value of the Higgsino parameter, hence of the Higgsino mass, saturating
the relic density constraint is smaller than for Winos. This is illustrated also
in the left-panel of Fig.~\ref{3Dlabels}, where iso-level curves for
$\Omega_{CDM} h^2 = 0.13$ are shown in the plane  $(M_2,\mu)$ and 
for a few values of $M_1$. In case $M_1$ is very heavy, we see the transition
of the LSP from being a pure Wino (upper part of the Figure) to being a pure
Higgsino (right-hand side). The size of the saturating parameters differs because the
coupling in the most important vertex, namely $W^+ \tilde{\chi}_1^0 \tilde{\chi}_i^+$, reads $g/\sqrt{2}$ for Higgsinos and  $g$ for Winos, but also because the 
number of degrees of freedom in the coannihilation process in the two cases 
is different: as far as Higgsinos are concerned, the coannihilating particles are two neutralinos and one chargino, all nearly
degenerate in mass,
while for Winos only one neutralino and one chargino enter (and this matters,
because pair annihilation rates of charginos are larger than those for 
neutralinos). We find that a pure Higgsino of mass equal to $1030$~GeV and
a pure Wino of mass $2210$~GeV have a relic abundance equal to
$\Omega_{CDM} h^2 = 0.113$, the current best fit value from the WMAP data.  

Lower values of $M_1$ imply smaller upper limits on the neutralino mass,
corresponding, as sketched before, to Bino LSPs  coannihilating either
with Winos or Higgsinos, with the transition regimes getting progressively wider
as $M_1$ decreases. A 3-D representation of the hypersurface at 
$\Omega_\chi h^2 =0.113$ in the $(M_1,M_2,\mu)$ space is given
in the right-panel of Fig.~\ref{3Dlabels}; the allowed low relic density region 
stands {\em behind} the plotted surface.

The value of $\tan\beta$ and $m_h$
have not played much of a role in our discussion, since we expect very minor 
changes in the cosmologically allowed portion of the parameter space if these
are varied (and we have indeed verified numerically that this is the case).
Such choice of parameters is actually crucial only for direct detection rates, 
as we will stress below.

Finally, the bounds we have derived are valid under the assumption of a 
standard cosmological setup, and including only thermal sources of 
dark matter. The bounds get tighter, or, from another perspectives, models 
that give here subdominant CDM candidates may become fully consistent
with observations, if one introduces particle physics models with non-thermal 
sources of CDM, such as neutralino productions from gravitinos or moduli decays~\cite{non-therm}, or considers cosmological scenarios with faster 
expansion rates at the time of decoupling: sample cases are cosmologies 
with a quintessence energy density term dominating at the LSP 
freeze-out temperature~\cite{quint,Profumo:2003hq,Profumo:2004ex}, 
anisotropic Universes with effective shear density terms~\cite{shear,Profumo:2004ex} 
or in scalar-tensor theories \cite{Catena:2004ba}. Relaxing the upper bound 
is also possible, though maybe slightly more contrived, as it would involve wiping out
(totally, or in part) the thermal relic component with an entropy injection:
a mechanism of reheating at low temperature, {\em i.e.} lower than the LSP 
freeze-out temperature which is in the range $m_\chi/20$ -- $m_\chi/25$, 
would give this effect~\cite{lowrehea}. However, excluding this latter possibility, 
it is interesting to notice that, in this generic Split SUSY setup, the cosmological
bound is indeed the only issue forcing one of the sectors of the
theory to be light, {\em i.e.} at a scale lower than 2.2~TeV.

\section{Dark Matter detection rates}
\label{sec:rates}

\begin{figure*}[!t]
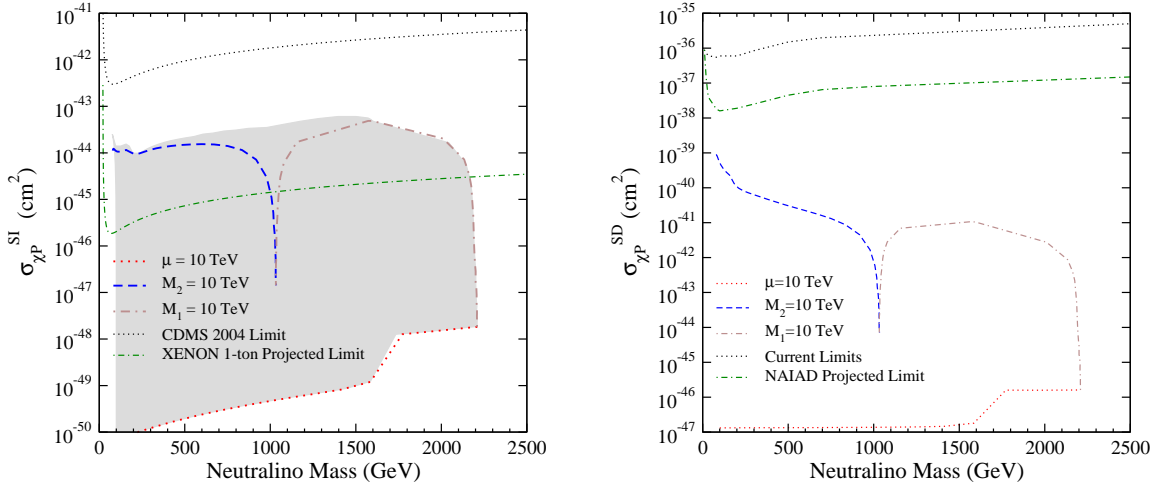

\begin{center}
\includegraphics[scale=0.5]{plots/CURR_Dirdet.eps}\qquad\includegraphics[scale=0.5]{plots/dirdetSD.eps}\\
\end{center}
\caption{\small \em Direct, spin-independent (left) and spin-dependent (right) neutralino-proton
scattering cross sections along the lines at
$\Omega_\chi h^2 =0.113$, obtained in the $(M_1,M_2,\mu)$ space
setting one of the parameters to 10 TeV and varying the other two.
The dotted black lines and the green dot-dashed lines correspond
to the current and projected experimental
limits. In the spin independent scattering cross sections plot, left panel, we also shade the region corresponding to the whole parameter space points giving the correct relic abundance, {\em i.e.} the hypersurface of Fig.~\ref{3Dlabels}, right panel.}\label{CURR_Dirdet}
\end{figure*}

The issue of WIMP dark matter detection has been studied at length 
(for a review, see, {\em e.g.},~\cite{Kamionkowski:1994dp,Bergstrom:1998xh};
see also the more recent works, e.g. Refs.~\cite{recwimpdet,baerdetect,Profumo:2004at}).
We will show here that this is especially relevant in the Split SUSY scenario
by systematically discussing all WIMP detection techniques that currently 
provide, or that will provide in the future, constraints on the model. 
Rates are computed with the \ds\ package~\cite{Gondolo:2004sc};
the set of underlining assumptions will be briefly described here. 
In this Section we will present sample rates for dark matter candidates in the 
three extreme regimes described above, and focusing on models
with thermal relic densities in the WMAP preferred range \cite{Spergel:2003cb},
in case of  a standard cosmological setup. In the next two Sections, 
these results will be generalized, including also low relic density models,
either by relaxing the requirement of thermal production of the whole CDM 
content of the Universe or by referring to non-standard cosmological scenarios;
at that stage we will also put more emphasis on the comparison
between different detection techniques.

In the last decade there has been a considerable experimental effort to
detect WIMPs directly, measuring the recoil energy from WIMPs elastic scattering 
on nuclei~\cite{dirdet}. Our theoretical predictions are derived with the 
usual effective Lagrangian approach, except that no contributions mediated by squarks
are present, and taking a standard set of parameters~\cite{Gasser,SMC} 
for nucleonic matrix elements (note, e.g., the strange content here is slightly 
smaller than the values implemented in other analyses, see~\cite{Gondolo:2004sc,paololars} 
for details).
In the left panel of Fig.~\ref{CURR_Dirdet}  we plot the 
spin-independent (SI) neutralino-proton scattering cross section 
$\sigma_{\chi P}^{\rm SI}$, as a function of the neutralino mass. 
The thick lines label models on the $\Omega_\chi h^2 =0.113$ isolevel curves 
in Fig.~\ref{m1m2} and \ref{3Dlabels}, left, i.e. the regimes with one fermion 
mass parameter heavier than the other two. From the upper left side of Figure, 
along a dashed line, we plot 
$\sigma_{\chi P}^{\rm SI}$ for a light LSP with significant Bino-Higgsino mixing
turning into a heavier pure Higgsino state; along the dashed-dotted curve the 
Higgsino state 
makes a transition into a pure Wino, going through models with large Higgsino-Wino
mixing, and finally on the dotted line the $\mu$ parameter is large and there
is just a very sharp transition from pure Winos to pure Binos when moving again to
smaller masses.  Since the SI scattering cross section is mediated by the light Higgs
in a t-channel, and the $\tilde{\chi}^0_1 \tilde{\chi}^0_1 h$ vertex scales with
the Higgsino-gaugino mixing, whenever this is small the cross section is small as 
well, as it can be clearly seen in the Figure.
The gray-shaded region corresponds to the projection of all points lying onto the hypersurface at $\Omega_\chi h^2=0.113$
in the $(M_1,M_2,\mu)$ space (the hypersurface plotted in Fig.~\ref{3Dlabels}, right).
Not surprisingly, this region is close to being the zone delimited by the curves labeling the three extreme regimes just described (and the same holds for all other detection techniques we discuss in the present Section, hence, for the sake of clarity, we omit the corresponding shadings). The black dotted line indicates the current bound on $\sigma_{\chi P}^{\rm SI}$ from the CDMS collaboration \cite{Akerib:2004fq}, while the green dot-dashed line stands for the projected sensitivity of next-generation (sometimes dubbed {\em stage 3}) detectors; for definiteness, we take as a benchmark experimental setup the XENON facility \cite{Aprile:2002ef}. As it can be seen, the present sensitivity
does not allow to set bounds on these models, but future projects will test a large portion 
of the parameter space.

In the right panel of Fig.~\ref{CURR_Dirdet} we show the spin-dependent neutralino-proton cross sections, along the same parameter space slices as in the left panel. We also indicate the current \cite{currsd} and projected \cite{naiad} experimental sensitivities in this detection channel. As already pointed out \cite{Profumo:2004at}, spin-dependent direct neutralino searches in the general MSSM appear to be largely disfavored with respect to spin-independent ones: future facilities will reach a sensitivity which is expected to be various orders of magnitude below the theoretical expectations for the models under consideration. 

\begin{figure*}[!t]
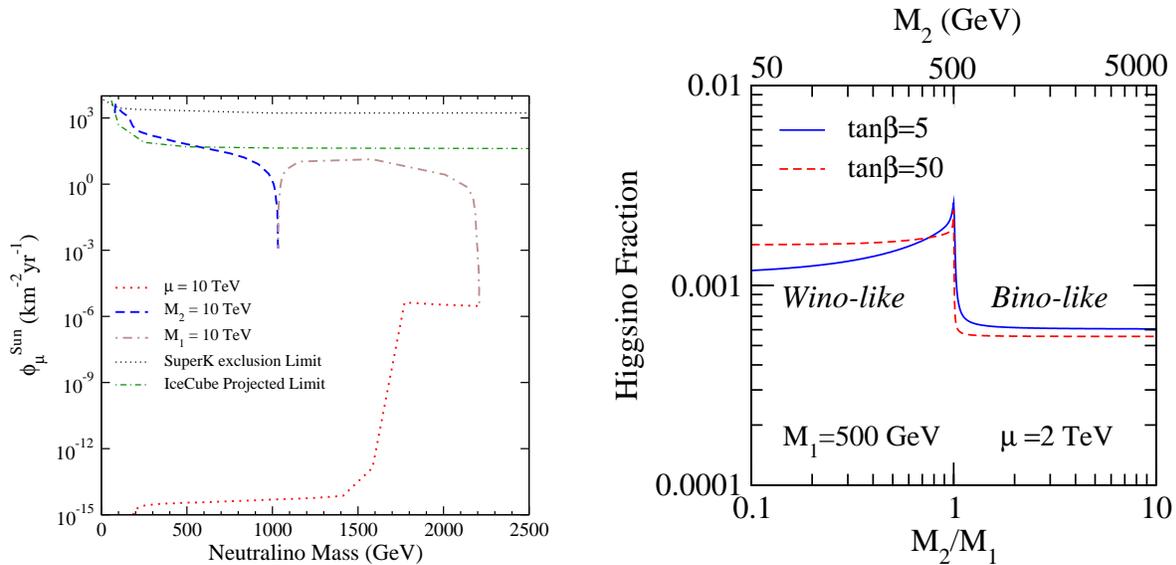

\begin{center}
\includegraphics[scale=0.5]{plots/CURR_Muflux.eps}\qquad\includegraphics[scale=0.7]{plots/higgsinofrac.eps}\\
\end{center}
\caption{\small \em Left: The neutralino-annihilations-induced muon flux from the Sun along the lines at
$\Omega_\chi h^2 =0.113$, obtained in the $(M_1,M_2,\mu)$ space
setting one of the parameters to 10 TeV, and varying the other two.
The dotted black lines and the green dot-dashed lines correspond
to the current and projected experimental
limits. Right: The Higgsino fraction, as a function of the parameter $M_2$, at
$M_1=500$ GeV and $\mu=2$ TeV, for two values of $\tan\beta$,
respectively 5 (solid blue line) and 50 (red dashed line).}\label{higgsinofrac}
\end{figure*}


The search for neutrinos produced by the annihilation of neutralinos trapped in the core of the gravitational wells of the Sun or of the Earth has been since long recognized as a promising indirect detection technique. In the present framework, fluxes
from the Earth are actually very low and will not be considered further;
to estimate neutrino fluxes from the Sun we implement the standard procedure 
described in Refs.~\cite{Bergstrom:1998xh,joakimnt},  except for a more careful
treatment of neutralino capture rates~\cite{sugrarates,Gondolo:2004sc}.
In the left panel of Fig.~\ref{higgsinofrac}, we present results
in terms of muon fluxes, above the threshold of 1~GeV, and compare them
to the current best limits from the SUPER-KAMIOKANDE Collaboration~\cite{Habig:2001ei} 
and with the future projected sensitivity of the
IceCube experiment \cite{icecube} (the mismatch in the energy threshold of
IceCube and the threshold considered here has been taken into account). The 
color coding on the $\Omega$ isolevel curves is the same as in the left
panel, making it transparent that the emerging pattern is perfectly specular.
The neutrino flux in the Sun scales with a capture rate which is dominated
by the spin dependent (SD) neutralino-proton coupling, and, as for the SI term,
this is suppressed in models with small Higgsino-Gaugino mixing (and the
relevant vertex here is $\tilde{\chi}^0_1 \tilde{\chi}^0_1 Z^0$).
While the sensitivity of current direct dark matter searches is more than one order of magnitude above the largest signals we obtain in the present framework, in the lowest mass range ($\sim100$~GeV) neutrino telescopes already rule out a few models. However, future direct detection experiments will probe a much wider portion of the parameter space, as compared to what IceCube is expected to achieve.

The results shown in Fig.~\ref{CURR_Dirdet} and \ref{higgsinofrac} are derived assuming, as in the previous Section, that $\tan\beta=50$ and $m_h = 115$~GeV. 
$\sigma_{\chi P}^{\rm SI}$ scales
with $m_h^{-4}$, while the dependence on $m_h$ of the neutrino fluxes from
the Sun is negligible. $\tan\beta$ does not enter critically in any of the relevant 
couplings, however it slightly affects the Higgsino-Gaugino mixing.  We sketch
a sample effect of this kind in the right panel of Fig.~\ref{higgsinofrac}, where the main point
is however to illustrate the
reason why the rates in the cases considered so far change sharply
at $m_\chi\simeq1.5\div 1.7$~TeV, in the Bino-Wino transition along the 
$\Omega$ isolevel curve at large $\mu$ (red dotted lines). The figure shows that
going from the regime\ $M_2<M_1$ to the regime $M_1<M_2$, there
is not only a sharp switch in the gaugino content, but also a step-like fall in the 
(already small) Higgsino fraction.  As we show, a slight dependence on $\tan\beta$ (which enters in the neutralino mass matrix) is present. We will further discuss 
this effect in the next Section.

Other indirect detection techniques rely on the fact that there is a finite 
probability for dark matter WIMPs to annihilate in pairs in galactic halos, and in 
particular in the halo of the Milky Way, possibly giving rise to exotic sources
of radiation and cosmic rays. In this analysis we will consider
the production of gamma-rays, antiprotons, positrons and antideuterons
within our own Galaxy.
The yield per annihilation in all these channels is read out of simulations 
with the {\tt{Pythia}} \cite{pythia} 6.154 Monte Carlo code as included
in the \ds package, except for $\bar{D}$, where we use the prescription 
suggested in Ref.~\cite{dbar}. Since the source strength  scales 
with the number density of neutralino pairs locally in space, i.e., in terms of 
the dark matter density profile, with $1/2\,(\rho_{\chi}(\vec{x}\,) / {m_{\chi}})^2$, 
the predictions for these signals are very sensitive to which $\rho_{\chi}(\vec{x}\,)$
is considered for the Milky Way. In fact, the latter is unfortunately not well-known,
and one is forced to make some assumptions.

In this analysis, we will mainly focus on two extreme models for the Milky Way
dark matter halo. They are both inspired to the current picture emerging
from N-body simulations of hierarchical clustering in $\Lambda$CDM 
cosmologies, respecting, e.g., the relation between halo mass and 
concentration parameter suggested therein. The two models diverge however
in the way the effect of the baryon infall in the dark matter potential well is
sketched. One extreme it to suppose that this process happened with a
large transfer of angular momentum between the baryons and the dark 
matter. In this case the system of cold particles, sitting at the
center of dark matter halo and forming the cusp seen in the simulations  
(which describe the evolution of the collision-less dark halo component only),
are significantly heated and removed from the central part of the halo. 
One such extreme scenario was proposed in Ref.~\cite{elzant}, suggesting 
that the resulting dark halo profile can be modeled by the so-called Burkert profile~\cite{burkert}:
\begin{equation}
  \rho_{B}(r) = \frac{\rho_B^0}{(1+r/a)\,(1+(r/a)^2)}\;.
\label{eq:burk}
\end{equation}
This profile has been shown to provide fair fits in case of a large sample of the rotation 
curves for spiral galaxies~\cite{salucci}, and, from our perspective, having
a very large core radius (the parameter $a$), is very conservative. We will consider
a sample choice of the free parameters in Eq.~\ref{eq:burk}, fixing the length scale parameter $a= 11.7$~kpc and the local halo density 
$\rho_B(r_0)= 0.34$~GeV~cm$^{-3}$. The second model we consider
stands on the other extreme, and it is derived by supposing that the settling in
of the baryons happened with no net transfer of angular momentum between baryons
and dark matter. The baryon infall happened as a slow and smooth process 
which gradually drove the potential well at the center of the Galaxy to become 
deeper and deeper, accreting more and more dark matter particles in this 
central region. This is the {\em adiabatic contraction limit}, which we implement according 
to the prescription of Blumenthal et al.~\cite{blumental} (circular orbit approximation). 
We start from the CDM profile of the non-singular form extrapolated in Ref.~\cite{n03}
(which we label as N03 profile), assume
as sample virial mass and concentration parameter, respectively, 
$M_{vir} = 1.8 \times 10^{12}\msun$ and $c_{vir} = 12$, derive the 
enhancement on it induced by the stellar bulge and the disc components, 
and just cut the profile in the region which is dynamically dominated by 
the central black hole (i.e. we assume complete relaxation due to the black
hole formation). The resulting spherical profile has a local dark matter
density equal to $\rho_{N03}(r_0)= 0.38$~GeV~cm$^{-3}$. The choice of
parameters in both scenarios is justified by testing the dark matter
model against available dynamical constraints. Finally, a self-consistent 
velocity distribution is derived as well, and the result for the Burkert profile
has already been exploited above when showing sensitivity curves for direct 
detection and computing capture rates in the Sun (and both of these change 
are essentially unchanged in case the adiabatically-contracted N03 profile is 
chosen). Further details on the definition of the halo models can be found in 
Refs.~\cite{sugrarates,Profumo:2004ty,halomod}.

\begin{figure*}[!t]
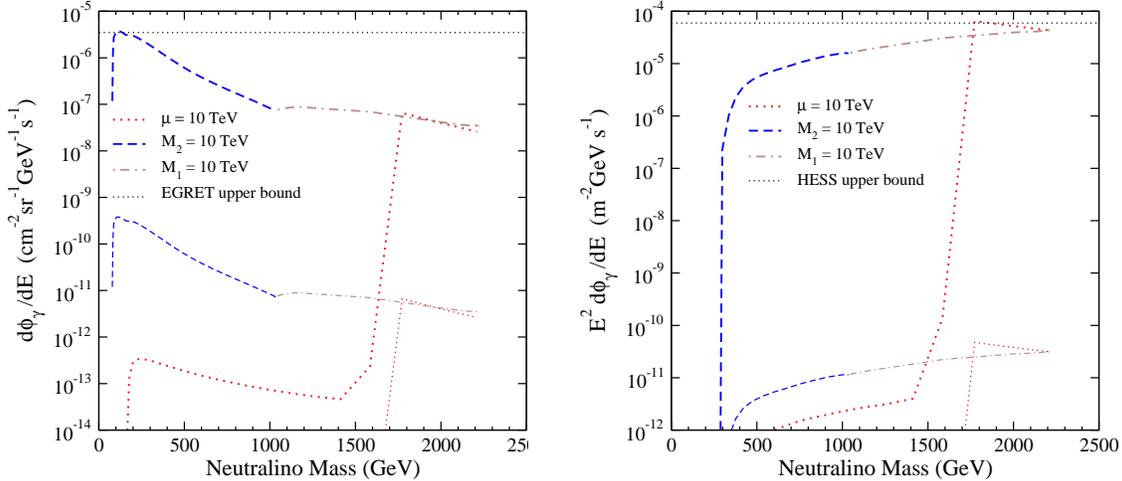

\begin{center}
\includegraphics[scale=0.5]{plots/CURR_Egret.eps}\quad\includegraphics[scale=0.5]{plots/CURR_Hess.eps}\\
\end{center}
\caption{\small \em Current constraints on the points at
$\Omega_\chi h^2 =0.113$ from gamma rays observations at the
center of our Galaxy, as performed by EGRET (left) and by H.E.S.S.
(right). The upper solid lines correspond to the N03-adiabatically
contracted halo model \cite{n03}, while the lower dotted lines to the
Burkert profile \cite{burkert}.}\label{CURR_Egret}
\end{figure*}

Antimatter searches do not play a main role for the models we have considered
so far, since we find that the predicted fluxes are well below both current 
constraints and the sensitivity of next-generation experiments; at the same time,
they look as the best chance for detection in other regions of the parameter space, 
hence we postpone the relative discussion. On the other hand,
data on the gamma-ray flux in the Galactic center (GC) direction are already
relevant to set combined constraints on our particle physics framework and the
dark halo profile. Gamma-ray maps of the GC have been derived both from space,
by the EGRET experiment on the Compton Gamma-Ray Observatory, below 
10~GeV~\cite{MH}, and, more recently, from ground by three different air 
Cherenkov telescopes (ACTs), with an overall energy coverage between 300~GeV
and 8~TeV, and apparently discrepant results concerning spectral shapes of the
source and normalization of the flux~\cite{whipple,cangaroo,Aharonian:2004wa}.
An interpretation of one of these signals in term of WIMP pair annihilations is possible,
see, e.g., the modeling of the EGRET source in Ref.~\cite{cesarini}, while 
most probably it is not possible to fit with such a source all of them simultaneously.
We take here a conservative attitude and just assume that the GC gamma-ray
flux produced in our model should not exceed the measured one; of course, in case
it were demonstrated that these fluxes were associated to another source, 
possibly not even located in the GC, as it has already been argued for
the EGRET source~\cite{hooper}, the constraints we derive here would get stronger.
A clearer statement on this respect should come with the next gamma-ray mission 
in space, the GLAST satellite \cite{glast}.

In Fig.~\ref{CURR_Egret}, left panel, we plot, with a black dotted horizontal line, the most stringent $2-\sigma$ bound from the EGRET data set~\cite{MH} (which turned out to be the bin at the largest energy $4\ {\rm GeV}<E_\gamma<10\ {\rm GeV}$), compared to the fluxes in this energy bin and computed for the cuspy, adiabatically contracted N03 model (upper lines) and cored Burkert profile (lower lines). The fluxes are averaged
over the EGRET angular resolution $\sim 1.5^{\circ}$. Going to larger masses the 
expected flux decreases both because of the $1/m_\chi^2$ scaling in the number 
density of neutralino pairs
and because we are looking at a rather low energy bin; the trend is partially 
smoothed, especially for Winos, by the transition along the $\Omega$ isolevel
curves into states that annihilate more and more efficiently into W bosons
(for pure Winos this is maximal). Note also, even in this case, the sharp drop
of the signal for pure Binos, which is now due to the fact that they can annihilate just
into fermions and with very suppressed rates (of course coannihilation effects make 
a compensation in the effective thermally averaged annihilation cross section setting 
the relic abundance in the early Universe, but not in the zero temperature cross 
section for annihilation for lightest state pairs, which is the relevant quantity
here). At higher energies we consider the constraint from the recent measurement
by the H.E.S.S. telescope, which has the largest statistics among recently 
reported results~\cite{Aharonian:2004wa}. Since the the flux detected by H.E.S.S. is 
surprisingly hard, the data point setting the strongest 
$2-\sigma$ bound is the one corresponding to the {\em smallest} energy bin, $E_\gamma=281$~GeV. The predicted fluxes at this energy are shown in
Fig.~\ref{CURR_Egret}, right panel; in this case they are averaged over a 
5.8$^\prime$ cone around the GC, corresponding to the H.E.S.S. angular 
resolution for this source. 
Obviously the fluxes are zero for neutralino masses below the energy we are 
considering; above that mass, the $1/m_\chi^2$ scaling in the number 
density of neutralino pairs is compensated (except for the pure Wino
branch in the heavy $\mu$ case) by the fact that we are considering a high energy
bin, in which the pile up of the produced photons get efficient just for heavy masses,
and again by the fact that the maximal pair annihilation rate involves pure Winos.
Note also the huge span in the predictions between the two halo models 
we are considering, even larger than in the case of the EGRET data, since the
angular acceptance here is smaller.

\begin{figure*}[!t]
\begin{center}
\includegraphics[scale=0.6]{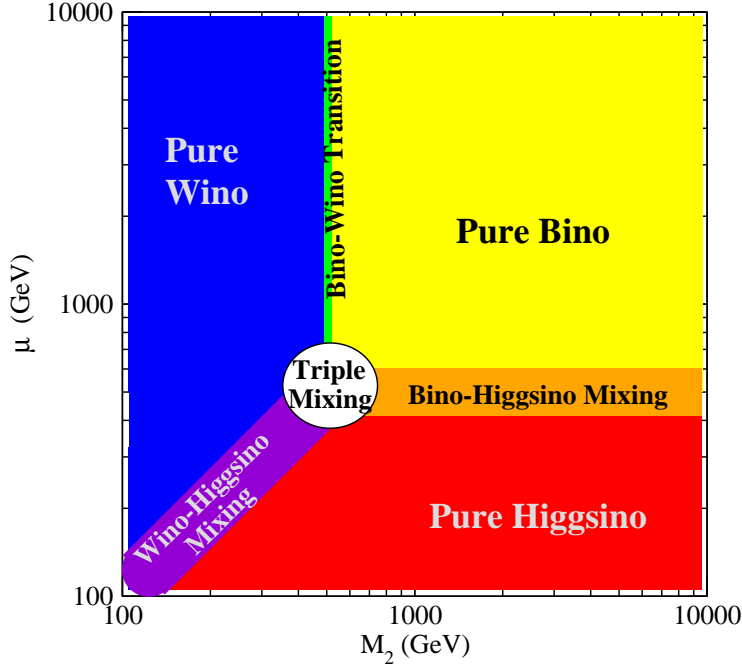}\\
\end{center}
\caption{\small \em A sketch of the neutralino composition in the case study plane $(M_2,\mu)$ at $M_1=500$ GeV. The plane collects a sample of all possible mixing patterns in the composition of the lightest neutralino within the MSSM.}\label{sketch}
\end{figure*}

\section{A case study}
\label{sec:casestudy}

\begin{figure*}[!t]
\hspace*{-1.5cm}\includegraphics[scale=1.05]{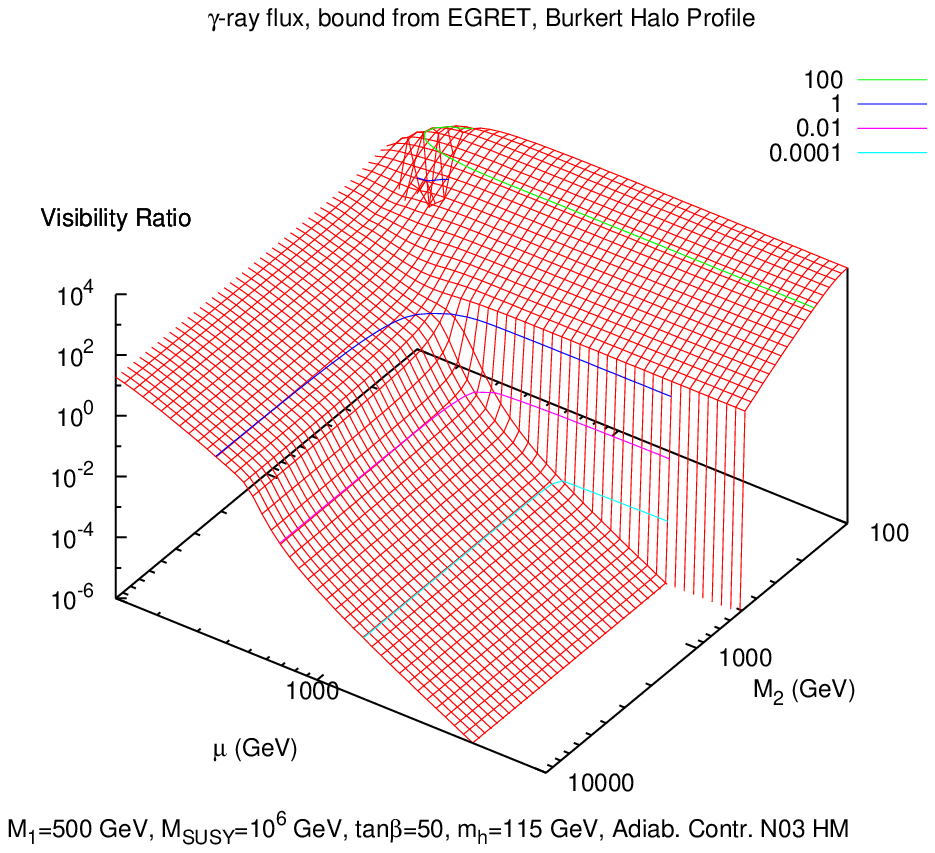}
\includegraphics[scale=1.0]{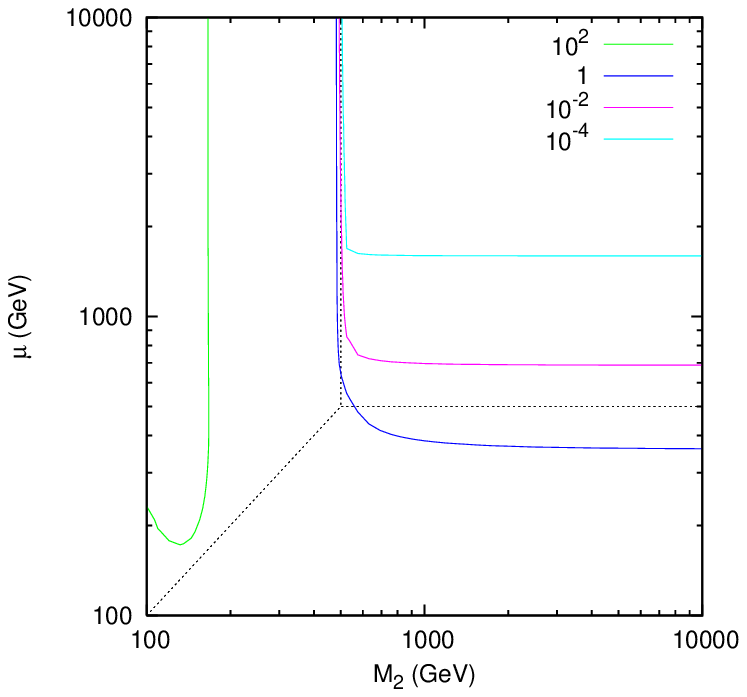}
\caption{\small \em The 95\% C.L. Exclusion Ratio from the EGRET
data (namely, the gamma ray flux from neutralino annihilation over
the maximal allowed observed gamma ray flux) on the $(M_2,\mu)$
plane at $M_1=500$ GeV. The N03-adiabatically contracted halo
model is assumed. In the right panel we show the corresponding
iso-level curves.}\label{Egret}
\end{figure*}

We are ready to examine in more detail the role of dark matter searches in the
Split SUSY framework. The approach we follow here is to choose a direction in the
parameter space, to postulate that these selected models provide the 
LSP as a dark matter candidate (i.e. relaxing the hypothesis that the
thermal relic abundance of $\chi$s in a standard cosmological setup should match
the WMAP preferred range) and to check their detectability.

In this Section we concentrate, if not differently specified, on a case study where we take the soft breaking mass of the hypercharge Gaugino $M_1=500$ GeV, $\tan\beta=50$ and $m_h=115$ GeV. This corresponds to a foliation of the parameter space along
$(\mu,\,M_2)$ planes; we consider a scan on these two parameters varying them in the range from 100 GeV to 10 TeV (we remind again that no GUT relation is assumed, and that $M_1$, $M_2$ and $\mu$ are considered as independent parameters). This sample case is particularly illuminating, because, within a simple and readable setting of the relevant low-energy parameters in the neutralino sector, it allows to investigate the full set of neutralino compositions. 
The pattern of mixing and composition for the LSP is schematically illustrated in 
Fig.~\ref{sketch}. We give here a detailed, and, to our knowledge novel, analysis of the physics occurring when the Wino and Higgsino mixing is large (violet strip in Fig.~\ref{sketch}), as well as an analysis of the Wino-Bino edge (green strip, in Fig.~\ref{sketch}).

\begin{figure*}[!t]
\hspace*{-1.5cm}\includegraphics[scale=1.05]{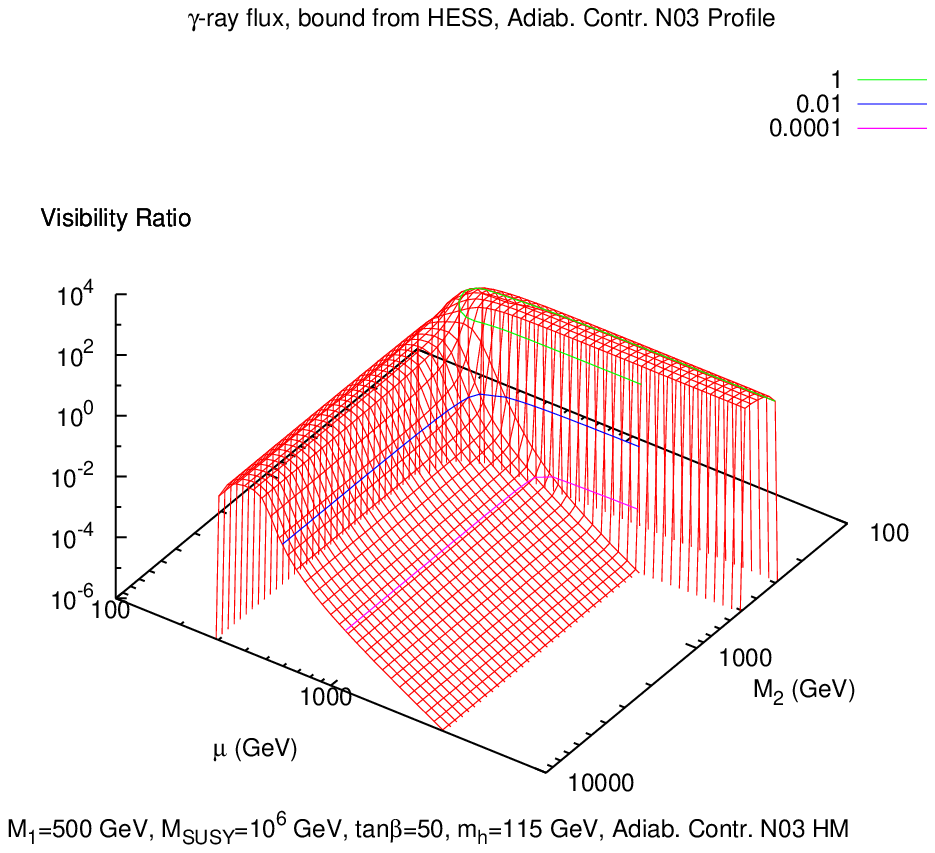}
\includegraphics[scale=1.0]{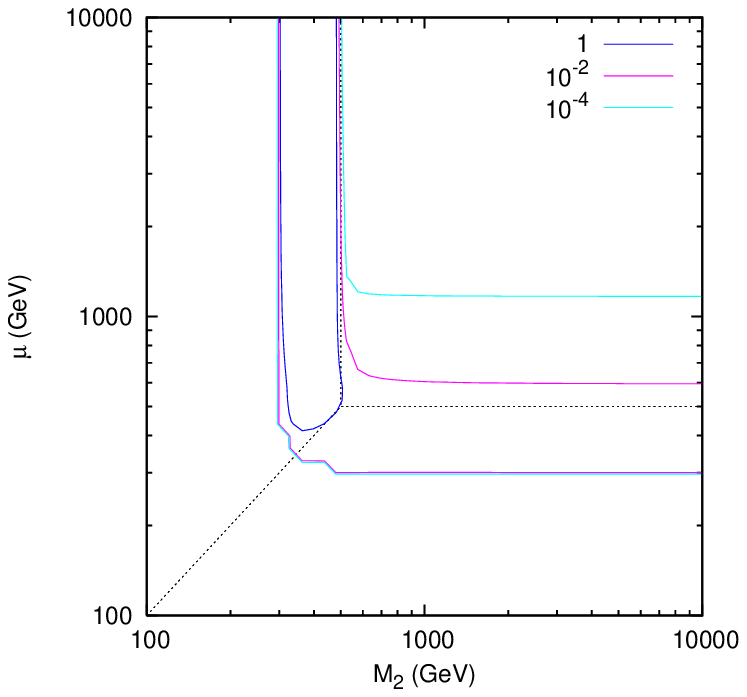}
\caption{\small \em The 95\% C.L. Exclusion Ratio from the H.E.S.S.
data on the $(M_2,\mu)$ plane at $M_1=500$ GeV, with the
N03-adiabatically contracted halo model. The right panel shows the
corresponding iso-level curves.}\label{Hess}
\end{figure*}

We present our results by resorting to {\em Visibility Ratios} (VRs), i.e. ratios of
expected signals to projected (or current) sensitivities.
A model featuring a VR$>1$ in a given search channel will therefore be within the projected reach of (or excluded by) the corresponding experimental facility.

We start with the existing bounds from measurements of the gamma-ray flux towards the GC. Fig.~\ref{Egret} shows the exclusion curves one draws from the EGRET results, under the hypothesis of subdominant background (see Sec.~\ref{sec:rates}). The VR
is defined as:
\begin{equation}
{\rm VR}_{\rm EGRET}\equiv
\left. \frac{\phi^\gamma_{\chi\chi}}{\phi^\gamma+2\sigma_\phi}\right|_{E_\gamma 
\in [4,10]~{\rm GeV}}\,.
\label{eq:egret}
\end{equation}
Here and in the analogous figures below, on the left we show 3D plots with values of 
VR on the $(M_2,\mu)$ plane, on the right the corresponding isolevel curves.
The cuspy adiabatically contracted N03 profile is assumed in the plots; fluxes and VRs get a factor of $10^4$ smaller in case of the cored Burkert halo.
The sharp decrease in the low masses regions, at $M_2\sim\mu\sim 100$ GeV, is due to the gauge boson threshold, {\em i.e.} when $m_\chi<m_W$, in which case the gamma rays production cross section suddenly decreases. We recall that this region is, however, already ruled out by chargino searches at LEP, see Fig.~\ref{3Dlabels}, right. The largest production of gamma rays occurs in the light $M_2$ region, and it does not depend either on $M_1$ or $\mu$. In the region of Bino like neutralino, the VRs sensibly decrease, along with the decrease in the Higgsino fraction, and the expected fluxes are always well below the measured signal. The EGRET data also constrain pure Higgsinos, up to masses around 350 GeV.

The VR from the H.E.S.S. measurement is defined analogously, except that now in 
Eq.~\ref{eq:egret} the reference energy bin is the lowest in the H.E.S.S. dataset, i.e the 
one at 281~GeV; this sets the threshold in VR at approximately 300 GeV. Once again, the largest signal is obtained for Wino like LSP's.  Relaxing the $M_1=500$ GeV assumption, we find that sufficiently pure Winos with masses up to around 2 TeV are
excluded by H.E.S.S. data, for the cuspy halo model we have considered
(but are perfectly viable in case of the cored profile, since fluxes are suppressed, in that case,
by more than six orders of magnitude). VRs for Higgsinos are smaller, but still many 
configurations are excluded, while again Binos give very small fluxes.

\begin{figure*}[!t]
\hspace*{-1.5cm}\includegraphics[scale=1.05]{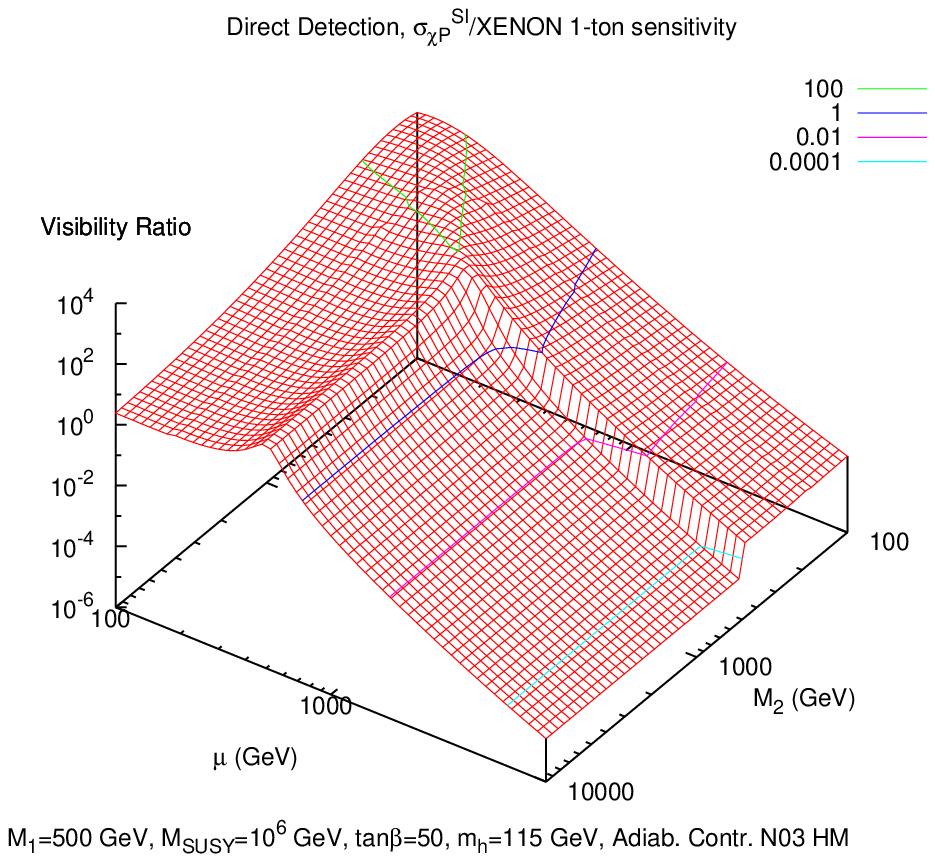}
\includegraphics[scale=0.95]{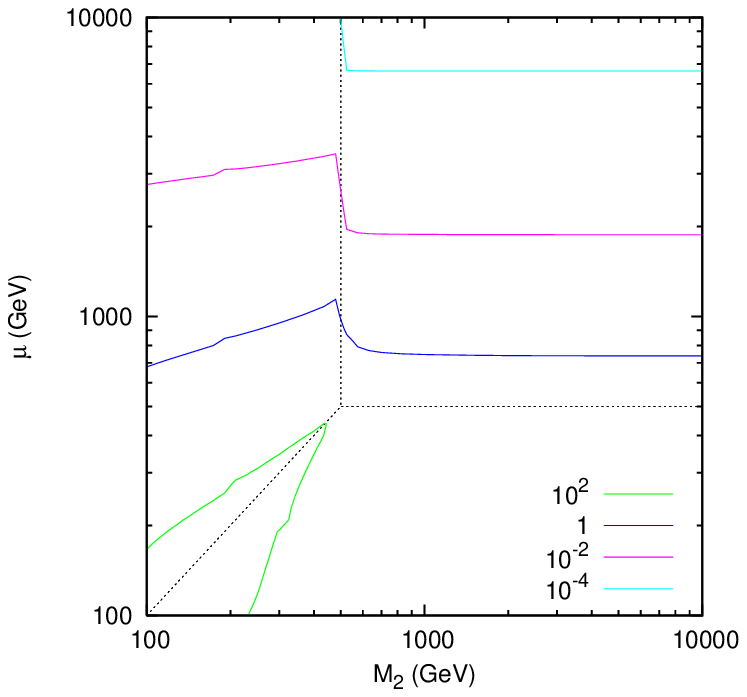}
\caption{\small \em The Visibility Ratio for Direct
spin-independent searches (expected signal over future projected
sensitivity for a XENON-1t like experiment) on the $(M_2,\mu)$
plane at $M_1=500$ GeV, and the corresponding iso-level
curves.}\label{Dirdet1}
\end{figure*}
\begin{figure*}[!h]
\hspace*{-1.5cm}\includegraphics[scale=0.9]{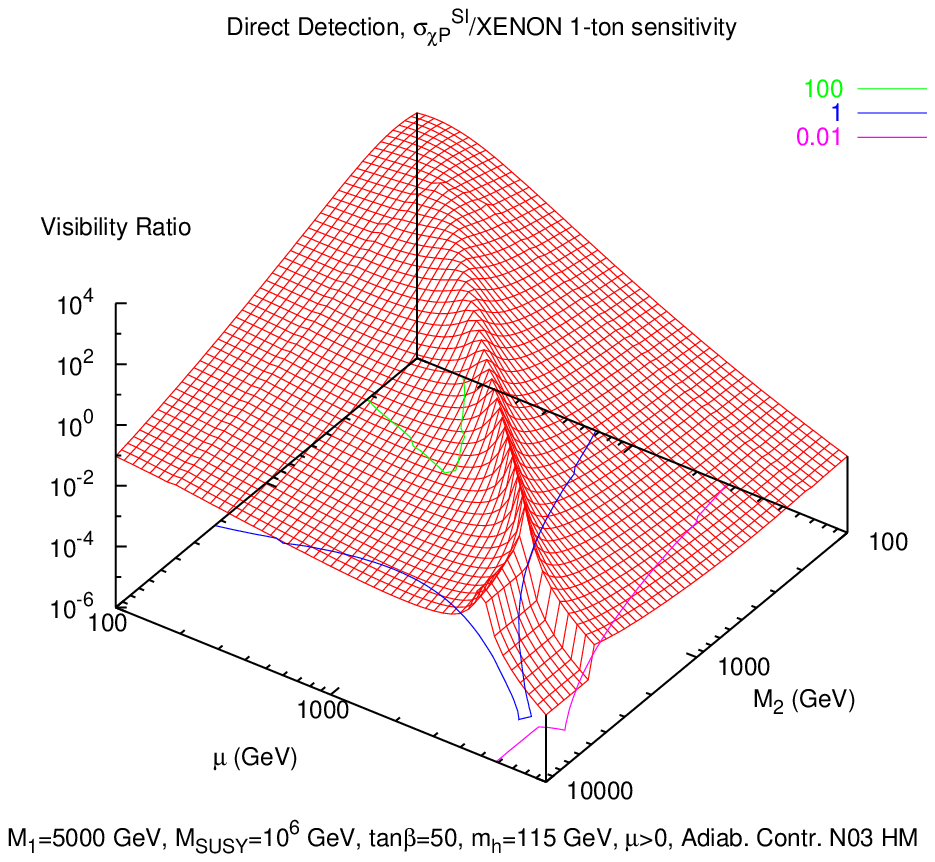}
\includegraphics[scale=0.9]{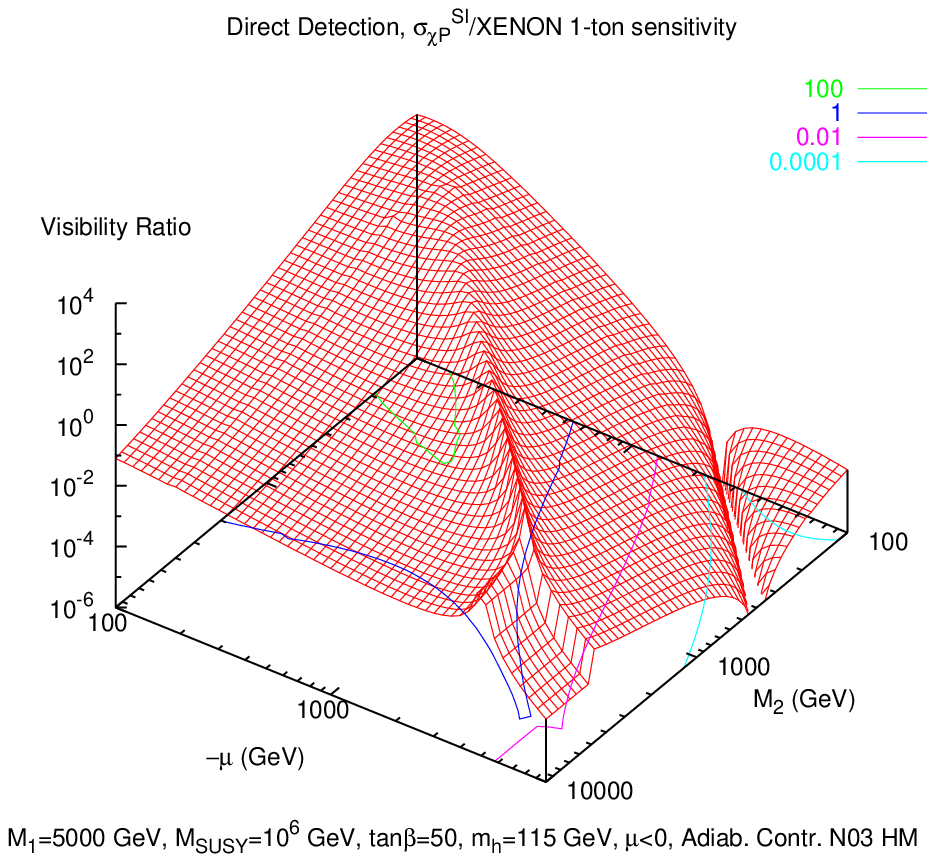}
\caption{\small \em The same as in Fig.~\ref{Dirdet1}, but with
$M_1=5000$ GeV, for $\mu>0$ (left) and $\mu<0$
(right). We draw the isolevel curves not on the surface but rather project them on the $(M_2,\mu)$ plane. Cancellations occur for $\mu<0$, since the Higgs-neutralino coupling $g_{h\chi\chi}\propto (\sin\alpha Z_{13}+\cos\alpha Z_{14})$; while for $\mu>0$ both terms have the same sign, at $\mu<0$ they have opposite signs, thus giving rise to the possibility of cancellations.}\label{Dirdet2}
\end{figure*}

Turning to spin-independent direct detection, we take here, as a benchmark of next generation search experiments, the future reach of the XENON-1ton facility \cite{Aprile:2002ef}. The corresponding projected sensitivity curve in the $(m_\chi,\sigma^{\rm SI}_{\chi P})$ is shown in the left frame of 
Fig.~\ref{CURR_Dirdet}; the VR is defined here as the ratio between the expected cross 
section and the projected sensitivity, at the corresponding neutralino mass.

\begin{figure*}[!t]
\hspace*{-1.5cm}\includegraphics[scale=1.05]{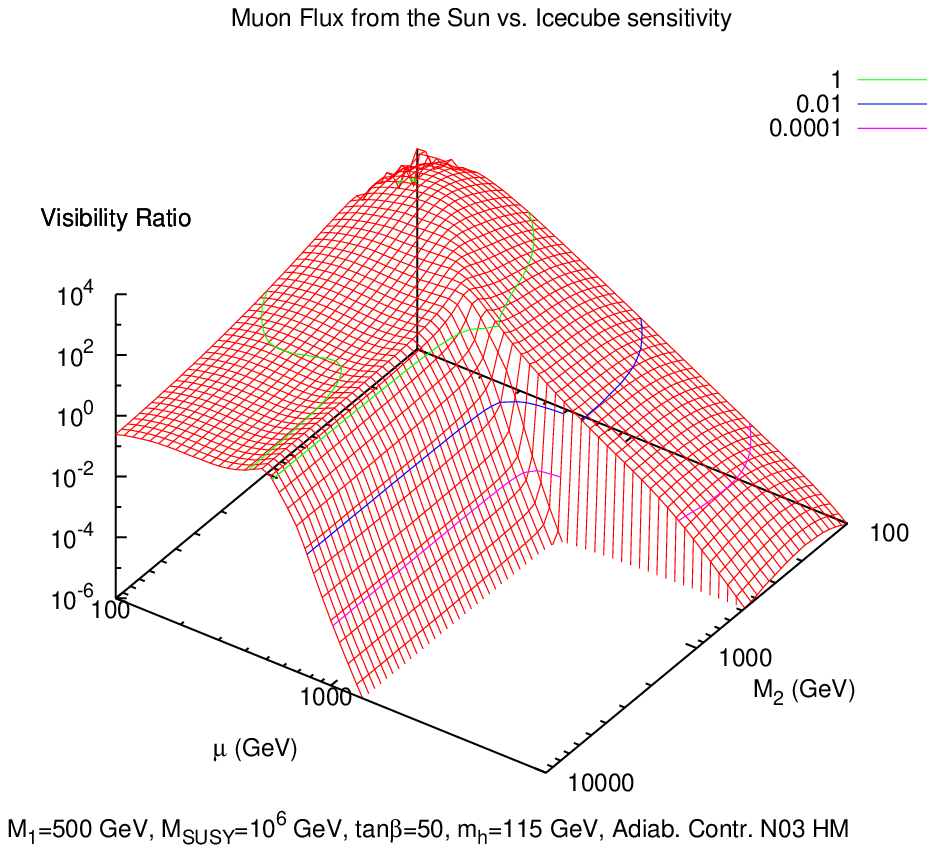}
\includegraphics[scale=0.95]{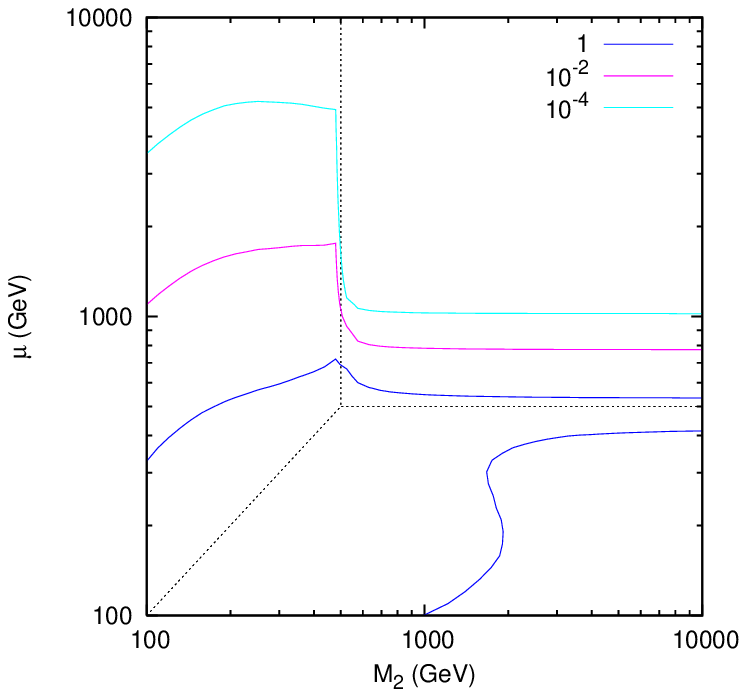}
\caption{\small \em The Visibility Ratio for the
neutralino-annihilations-induced muon flux from the Sun (expected
signal over future projected sensitivity for the IceCube
experiment) on the $(M_2,\mu)$ plane at $M_1=500$ GeV, and the
corresponding iso-level curves.}\label{Muflux}
\end{figure*}

Fig.~\ref{Dirdet1} highlights again the role of Gaugino-Higgsino mixing in direct detections searches: $\sigma^{\rm SI}_{\chi P}\propto (Z_{h}Z_g)^2$, with $Z_h$ 
and $Z_g$, respectively, the lightest neutralino Higgsino and Gaugino fractions;
their product is maximized on the ridge along the $M_2\simeq\mu$ line (explaining the shape of the VR=100 isolevel curve) and has a line of local maxima around 
$\mu\simeq M_1=500$~GeV. 
Switching from a Bino like to a Wino like lightest neutralino, {\em e.g.} on lines at constant $\mu>500$ GeV, the scattering cross section is enhanced by one order of magnitude or even more. This can be retraced to the sudden increase in the Higgsino fraction when the neutralino becomes Wino like we have shown in the right panel of Fig.~\ref{higgsinofrac}, and 
 slightly depends on $\tan\beta$. The region at $\mu,M_2>M_1=500$ GeV features a 500 GeV Bino like neutralino, whose scattering cross section off a proton keeps decreasing as the Higgsino fraction, with increasing $\mu$.
The isolevel curves indicate that, in this case study, neutralinos as heavy as 700-800 GeV will be within reach of future detection facilities; the critical parameter is found to be the value of $\mu$, while the $M_2$ dependence is quite mild. 

\begin{figure*}[!t]
\hspace*{-1.5cm}\includegraphics[scale=1.05]{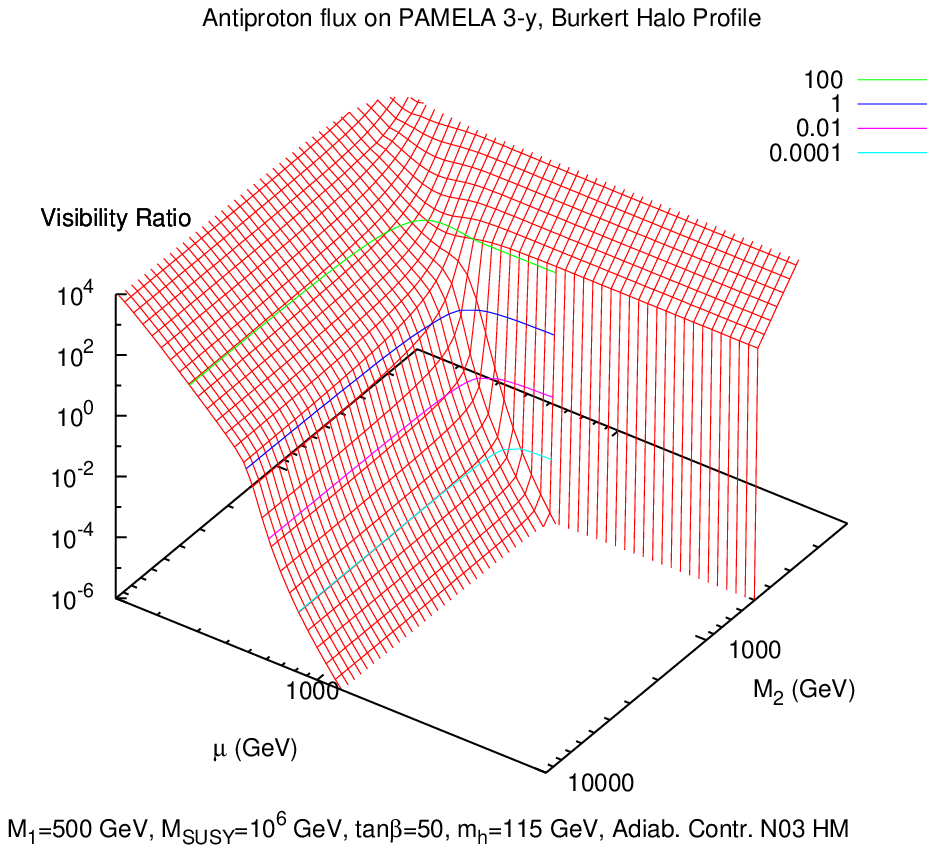}
\includegraphics[scale=0.95]{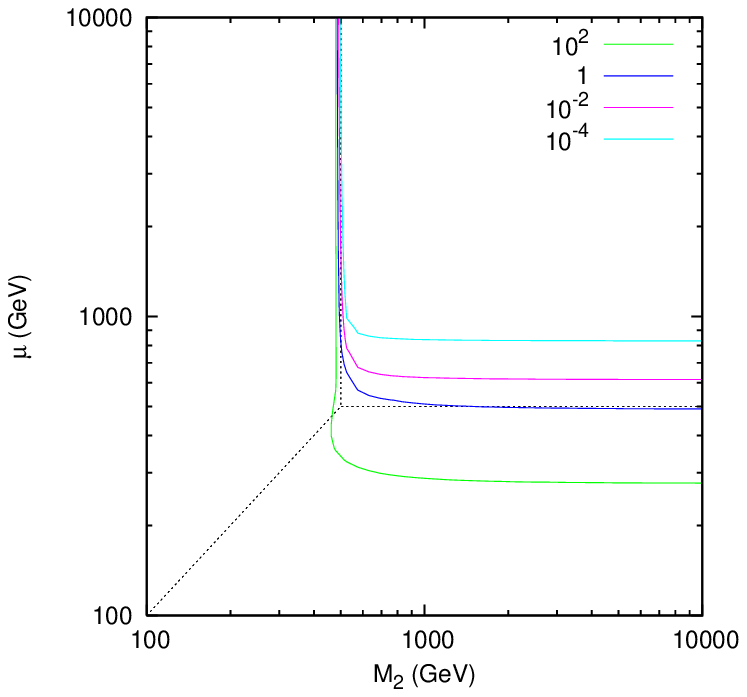}
\caption{\small \em The Visibility Ratio for antiprotons searches
(expected value of the $I_\phi$ parameter over future
corresponding projected sensitivity for the Pamela experiment
after 3 years of data-taking) on the $(M_2,\mu)$ plane at
$M_1=500$ GeV, and the corresponding iso-level
curves.}
\label{fig:pbar}
\end{figure*}

We show in Fig.~\ref{Dirdet2} the VR for a larger value of $M_1=5$ TeV, and for both signs of $\mu$. In this case, the reach of direct searches is clearly correlated to the degree of Wino-Higgsino mixing. Strikingly enough, we find that, for both signs of $\mu$, neutralinos as heavy as 5 to 6 TeV will give a detectable signal at next generation direct detection searches, provided $M_2\sim\mu$. We emphasize that this situation is not theoretically unrealistic, as it is naturally realized along the hyperbolic-branch / focus-point (HB/FP) of mSUGRA (see ref.~\cite{}, where the phenomenology of bino-higgsino neutralino dark matter has also been extensively addressed); these is the slice in the parameter space close to the region where there is no viable electroweak symmetry breaking) of models featuring a Wino like LSP, as in the mAMSB model~\cite{amsb}. Interestingly enough, we find cancellations in $\sigma^{\rm SI}_{\chi P}$ for $\mu<0$ (see fig.~\ref{Dirdet2}, right). While in the past it was realized that interferences of heavy and light Higgs bosons exchanges could induce low values for the scattering cross sections, we point out here that what we find is of a different nature. In Split SUSY, in fact, the heavy Higgs contribution vanishes, and the $\sigma^{\rm SI}_{\chi P}$ is driven to low values by the neutralino-neutralino-Higgs coupling itself. The latter is in fact $g_{h\chi\chi}\propto (\sin\alpha Z_{13}+\cos\alpha Z_{14})$, hence a suitable interplay of $M_2$ and $\mu$ may entail, for a given value of $\tan\beta$, $g_{h\chi\chi}\simeq0$. Apart from this novel feature, we remark that the sign of $\mu$ does not affect much the VR in this channel. The same conclusion holds for all other detection channels: this is why we always restrict to the case $\mu>0$.

\begin{figure*}[!t]
\hspace*{-1.5cm}\includegraphics[scale=1.05]{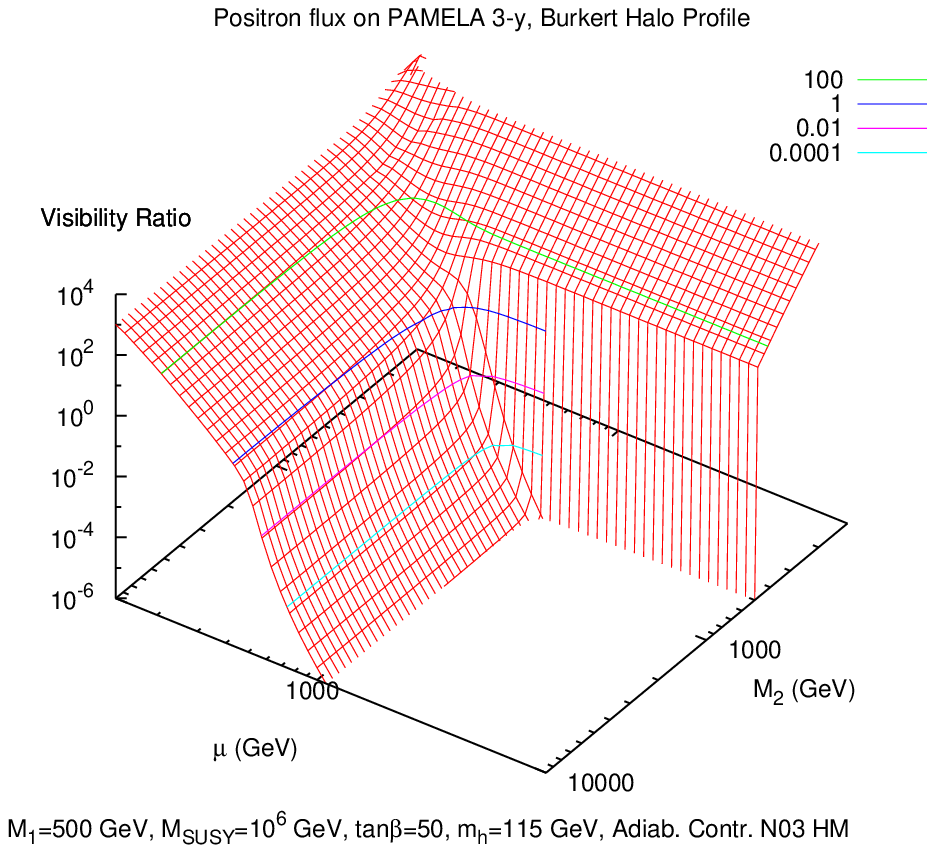}
\includegraphics[scale=0.95]{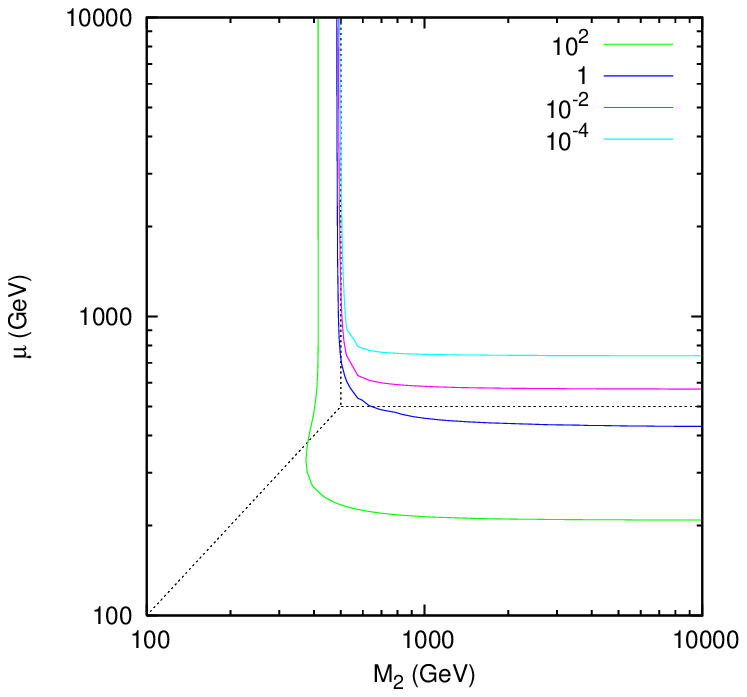}
\caption{\small \em The Visibility Ratio for positrons searches
(expected value of the $I_\phi$ parameter over future
corresponding projected sensitivity for the Pamela experiment
after 3 years of data-taking) on the $(M_2,\mu)$ plane at
$M_1=500$ GeV, and the corresponding iso-level
curves.}
\label{fig:eplus}
\end{figure*}

To examine the prospects for indirect detection with neutrino telescopes, we consider as a projected future sensitivity the IceCube expected limit on the muon flux from the Sun \cite{icecube}, {\em i.e.} the green curve of Fig.~\ref{CURR_Dirdet}, right. The intensity of the signal we find highlights again the role of Wino-Higgsino mixing: the line at $M_2\simeq\mu$ produces in fact the largest muon fluxes, as it combines large capture rates in the Sun, due to large spin-dependent neutralino-matter couplings, and large annihilation rates. As the LSP turns into a Bino, the capture-annihilation equilibrium ceases, and the VR suddenly collapses. In the Wino like regions, at $M_2<M_1,\mu$, we observe the role of the Higgsino fraction at fixed neutralino mass ({\em i.e.} along a given constant $M_2$), an effect which is fully accounted for by the decrease in the capture rate of neutralinos inside the Sun. The same effect is visible, though less evident, in the Higgsino like region.

Turning to antiprotons and positrons searches, we address the perspectives of discrimination of the signal from neutralino pair annihilations against the background
from the interaction of primary cosmic ray with the interstellar medium, 
following the approach outlined in Ref.~\cite{Profumo:2004ty}. We define the quantity:
\begin{equation}
I_\phi\equiv\ \int_{E_{\rm min}}^{E_{\rm max}} \frac{\left[\phi_s(E)\right]^2}
{\phi_b(E)}{\rm d}E\,,
\end{equation}
where $\phi_s(E)$ and $\phi_b(E)$ refer, respectively, to the signal and the background flux, and the integral extends over the energy range in which the integrand is 
non-negligible.  This quantity stems from the continuum limit, up to an overall
factor accounting for the  exposure times effective area of a given future experiment, 
of a $\chi^2$ statistical variable defined assuming that the background is known and the 
signal is subdominant with respect to the background. Antiproton and positron backgrounds are obtained with the \code{Galprop}~\cite{galprop} propagation code,
in a scheme with diffusion and convection tuned to give a fair estimate of ratios of primary to secondary cosmic ray nuclei. Signals are obtained within the same 
propagation framework. VRs are defined as ratios between $I_\phi$ and the
critical value in this quantity as estimated for the PAMELA 
experiment~\cite{pamela} after three years of data taking, i.e. 
$I_\phi^{3{\rm y}}=3.2\cdot 10^{-8}\, \, {\rm cm}^{-2}{\rm sr}^{-1}{\rm s}^{-1}$, see Ref.~\cite{Profumo:2004ty} for details. 

In Figs.~\ref{fig:pbar} and~\ref{fig:eplus} we plot visibility ratios in case of the adiabatically contracted N03 halo model; contrary to gamma rays, since fluxes are dominated by nearby sources, the suppression for the cored profile is not dramatic and 
discrimination pattern are comparable as we will discuss in Sec.~\ref{sec:discussion}.
Not surprisingly, we find a large dependence on the neutralino mass, and on the Higgsino fraction in the Bino like neutralino region at $M_2,\mu>M_1$. While antiprotons fluxes will be sensitive to Winos and Higgsinos up to masses as large as 500 GeV, the sensitivity in the positrons fluxes is somewhat less stringent, extending up to Higgsinos as heavy as 400 GeV. The fact that annihilating neutralinos give larger signal to background ratios for antiprotons than for positrons, has been already pointed out \cite{Profumo:2004ty,Profumo:2004at}, and we give here a further confirmation to this point.

\begin{figure*}[!t]
\hspace*{-1.5cm}\includegraphics[scale=1.05]{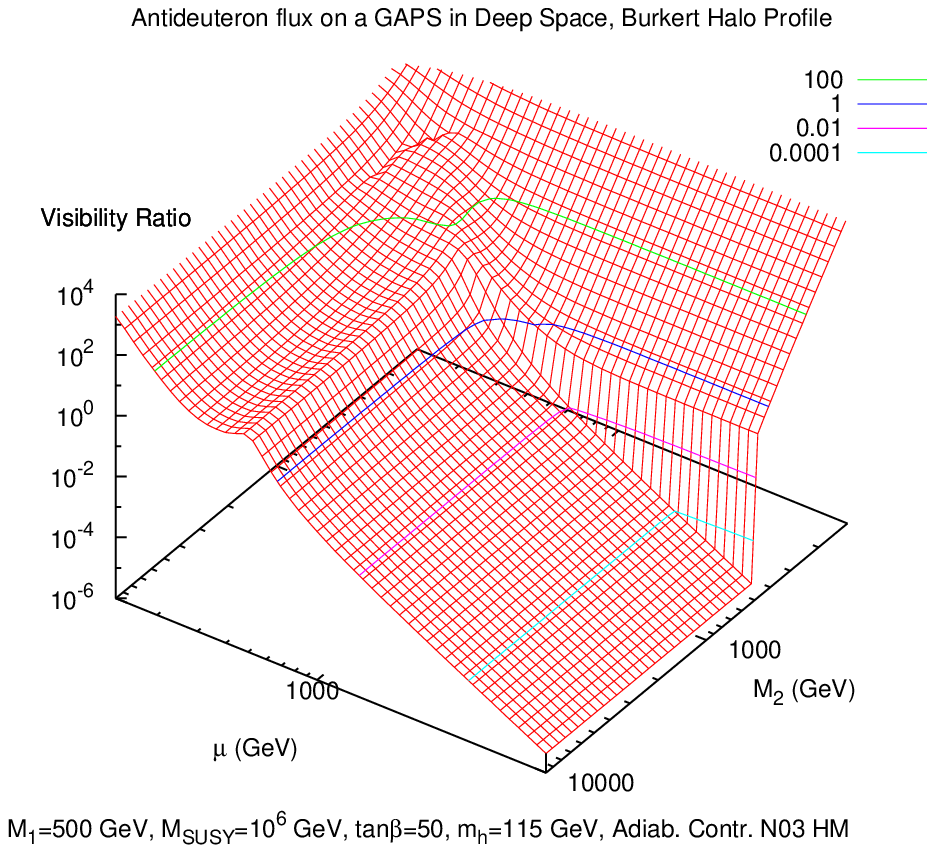}
\includegraphics[scale=0.95]{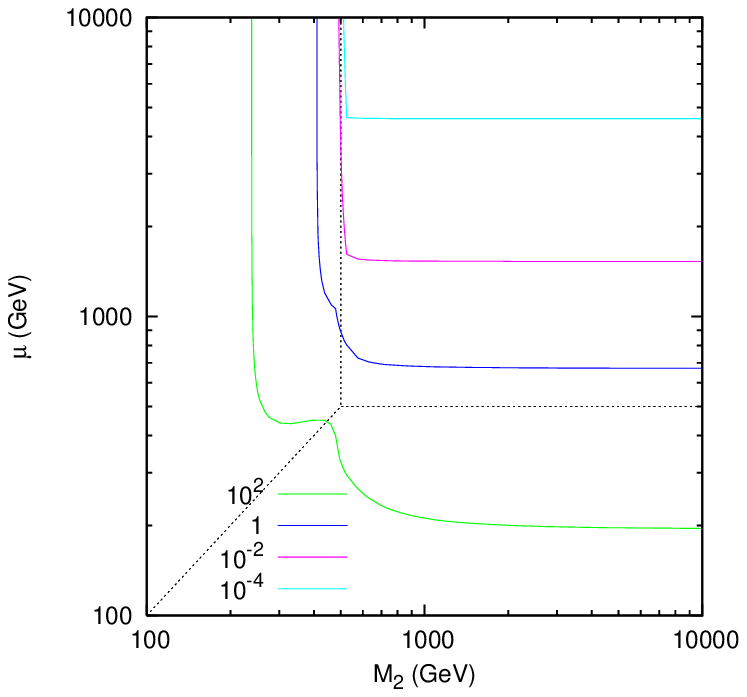}
\caption{\small \em The Visibility Ratio for antideuterons
searches (expected signal over the projected sensitivity for a
GAPS-like experiment) on the $(M_2,\mu)$ plane at $M_1=500$ GeV,
and the corresponding iso-level curves.}\label{Antideuteron}
\end{figure*}

The case for antideuterons is different because here, restricting to
a low energy window, the background flux is expected to be negligible~\cite{dbar},
and even detection of 1 event would imply discovery of an exotic component. 
Regarding the detection prospects, we will consider,
as the ultimate reach for an experiment in the future, that of the gaseous antiparticle 
spectrometer (GAPS) \cite{Mori:2001dv}. This is a proposal for an instrument
looking for antideuterons in the energy interval 0.1-0.4 GeV per nucleon, with
estimated sensitivity level of $2.6\times10^{-9}\textrm{m}^{-2}\textrm{sr}^{-1}\textrm{GeV}^{-1}\textrm{s}^{-1}$, to be placed either on a satellite orbiting
around the earth or on a probe to be sent into deep space. We compute the antideuteron flux induced by neutralino annihilations in this energy bin, and define the VR as the ratio between this and the mentioned expected sensitivity of GAPS~\cite{Profumo:2004ty}. 

Results for VRs in Fig.~\ref{Antideuteron} show that a critical threshold in the signal is that of top quarks; once again, the signal is maximized along the largest Wino-Higgsino mixing line. The Bino-Higgsino mixing effect is also clearly distinguishable in the line of maxima in VR around $\mu\simeq M_1<M_2$; the Higgsino fraction is found to enhance the antideuteron production (region $\mu> M_1\ll M_2$). As discussed in \cite{Profumo:2004ty}, Binos are somewhat less disfavored in antideuterons rather than in antiprotons or positrons searches, due to the hadronization of Bino's dominant decay products, such as $b\bar b$. The critical detection line (VR=1) extends here well beyond that of future antiprotons and positrons searches, as envisaged in Ref.~\cite{Profumo:2004ty}, entering the region of Bino like neutralinos with masses as large as 700 GeV.

\begin{figure*}[!t]
\hspace*{-1.5cm}\includegraphics[scale=1.05]{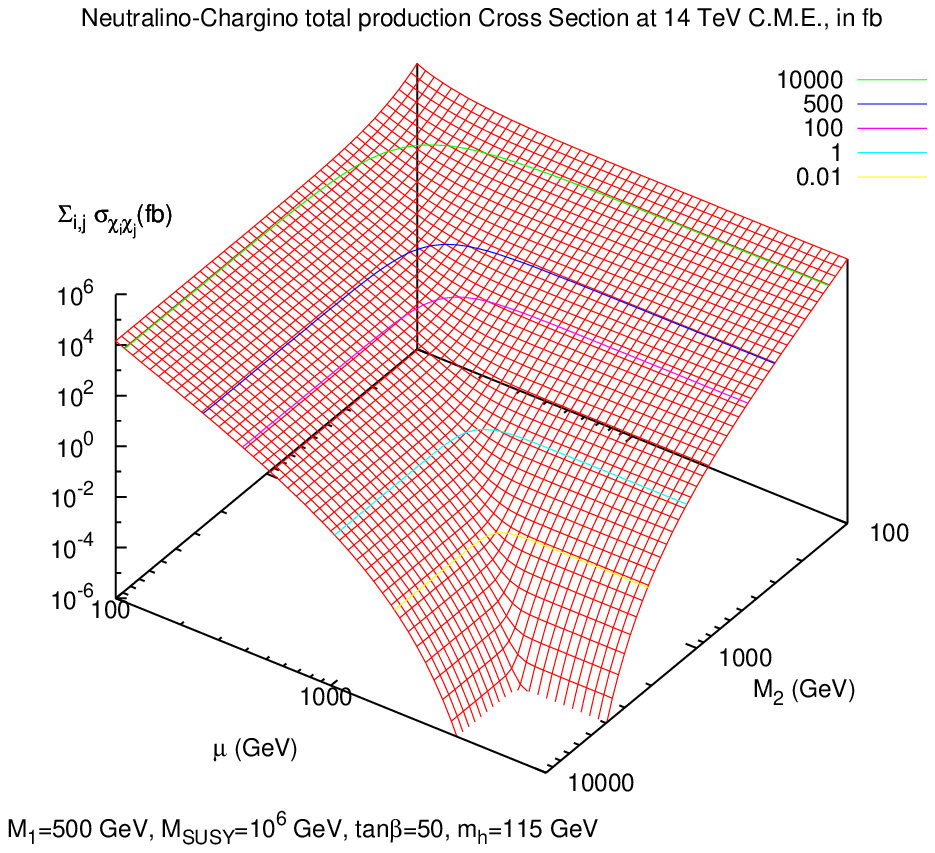}
\includegraphics[scale=0.95]{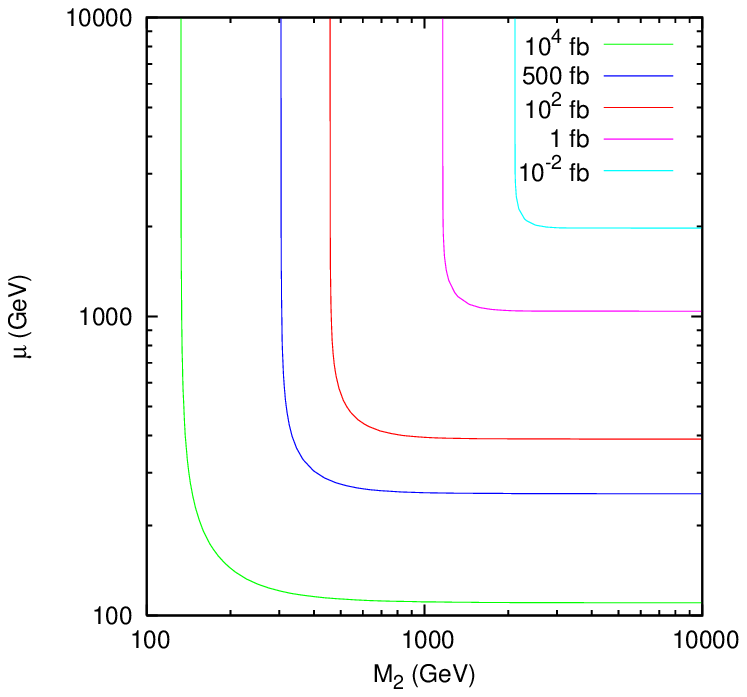}
\caption{\small \em The total production cross section for all
possible neutralino-neutralino, chargino-neutralino and
chargino-chargino reactions at the LHC (center-of-mass energy of
14 TeV), in fb, and the corresponding iso-level
curves.}\label{LHC}
\end{figure*}

The last process we consider is the production of neutralinos and charginos at the CERN Large Hadron Collider;  we computed, with the {\tt Prospino2.0} package \cite{Beenakker:1996ed}, the total production cross section of charginos and neutralinos at a center of mass energy of 14 TeV, {\em i.e.}:
\begin{equation}
\sigma_{\rm TOT}\equiv\sum_{i<j=1}^4 \sigma_{\chi^0_i\chi^0_j}+\sum_{i,j=1}^{4,2} \sigma_{\chi^0_i\chi^\pm_j}+\sum_{i,j=1}^2 \sigma_{\chi^\pm_i\chi^\mp_j}.
\end{equation}
A thorough analysis of the expected LHC background and of the suitable cuts to optimize searches for light neutralinos and charginos goes beyond the scope of the present work. We outline, however, that, in two extreme cases, dedicated studies have faced this very same problem: pure Winos in mAMSB with large common scalar masses $m_0$, as discussed in \cite{Baer:2000bs} and more recently in \cite{Barr:2002ex}; and Higgsinos in the HB/FP region of mSUGRA, which have been addressed in \cite{Baer:2003dj,Baer:2003wx}. We find that in both cases the critical total production cross section tentatively amounts to values around 500fb, which we reproduce for illustrative purposes in the iso-level curves.

\subsection{Summary of the case study results}

\begin{figure*}[!t]
\begin{center}
\includegraphics[scale=0.52]{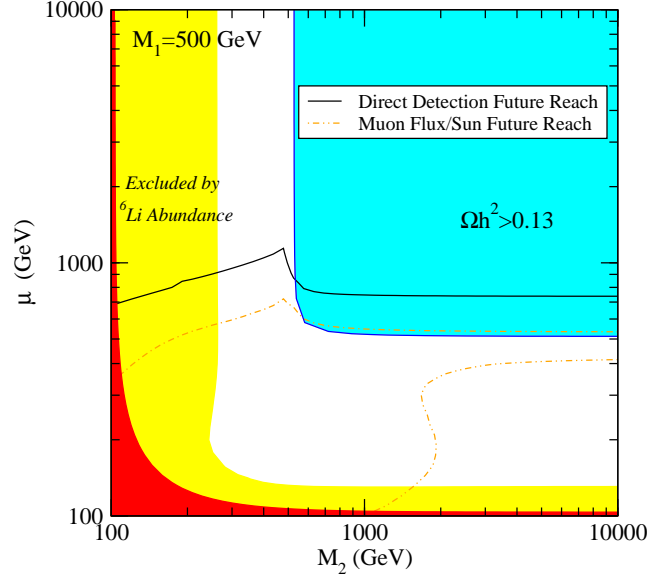}\\
\end{center}
\caption{\small \em The reach of future facilities in the
direct detection and muon-flux from the Sun channels, on the
$(M_2,\mu)$ plane at $M_1=500$ GeV. We shade in yellow the region which is not consistent with the measured Lithium 6 abundance.}\label{summary1}
\end{figure*}

\begin{figure*}[!t]
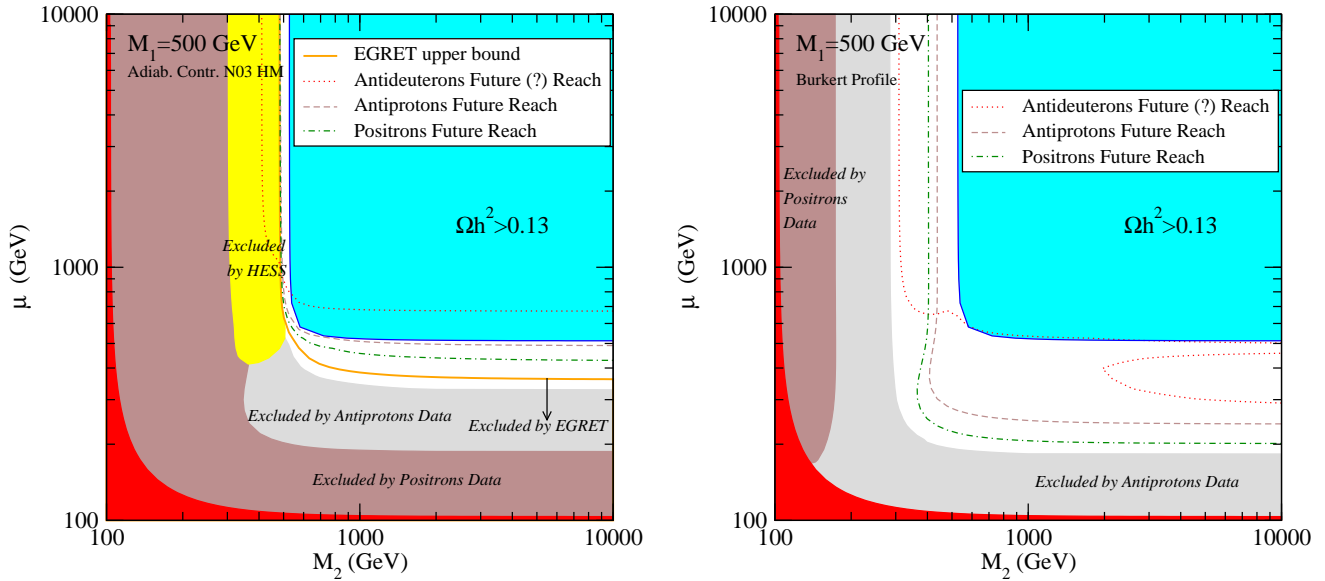

\vspace*{0.5cm}\hspace*{-0.5cm}\includegraphics[scale=0.52]{plots/summary500_2.eps}\quad\includegraphics[scale=0.52]{plots/summary500_3.eps}\\
\caption{\small \em The reach of future facilities for neutralino
detection through antimatter searches on the $(M_2,\mu)$ plane at
$M_1=500$ GeV, with the N03-adiabatically contracted profile
(left) and with the Burkert profile (right). The gray and brown shadings respectively reproduce the parameter space inconsistent with current data on Antiprotons and Positrons. We also indicate the
EGRET exclusion bound and shade the region excluded by the H.E.S.S.
data (these constraints are null for the Burkert cored
profile).}\label{summary2}
\end{figure*}

With the aim of comparing different future neutralino dark matter detection strategies in the case study under scrutiny, we collect in Fig.~\ref{summary1} and \ref{summary2} the iso-level curves corresponding to the critical VR=1 lines. We also include the bound stemming from the primordial production of ${}^6$Li~\cite{Jedamzik:2004er}
induced by the residual pair annihilations of neutralinos, after freeze-out and 
during the period of synthesis of light elements (the production of of ${}^6$Li
in the standard Bing Bang nucleosynthesis scenario is very strongly suppressed).
The limit we implemented is extrapolated from ref.~\cite{Jedamzik:2004er}
for the cases of interest here (namely those of neutralino annihilations with gauge 
bosons final states); we shade the corresponding excluded regions in Fig.~\ref{summary1}, where we show also the reach contours of direct spin-independent detection and of the muon-flux from the Sun. Fig.~\ref{summary2} shows our results concerning halo-dependent quantities: results for the cuspy adiabatically contracted N03 profile and for the Burkert cored profile are respectively given in the left and right panels. 

Interestingly enough, we find that the region excluded by the ${}^6$Li bound rules out a wide portion of parameter space where direct searches are not effective, namely that of pure Winos. We also remark that the whole region covered by the detection of a neutralino induced muon flux from the Sun is completely contained in that of direct searches: a positive sign from IceCube would imply, within this framework, a visible signal at next generation direct spin-independent facilities. 

Fig.~\ref{summary2} shows the excluded regions from current data on antimatter fluxes~\cite{bess,capricepbar,heat,capriceeplus}: remarkably, in both halo models a large portion of pure Winos and Higgsinos is already ruled out by available data. Constraints from gamma rays are effective only in the case of a cuspy profile, and exclude an even larger portion of models. Future antimatter searches will cover essentially the whole parameter space in the large $\mu$ region; in the Higgsino-Bino mixing region, on the other hand, a clear hierarchy among antideuterons, antiprotons and positrons is visible; with a cuspy profile, the region will be in all cases thoroughly probed by the Pamela experiment.

\begin{figure*}[!t]
\begin{center}
\includegraphics[scale=0.6]{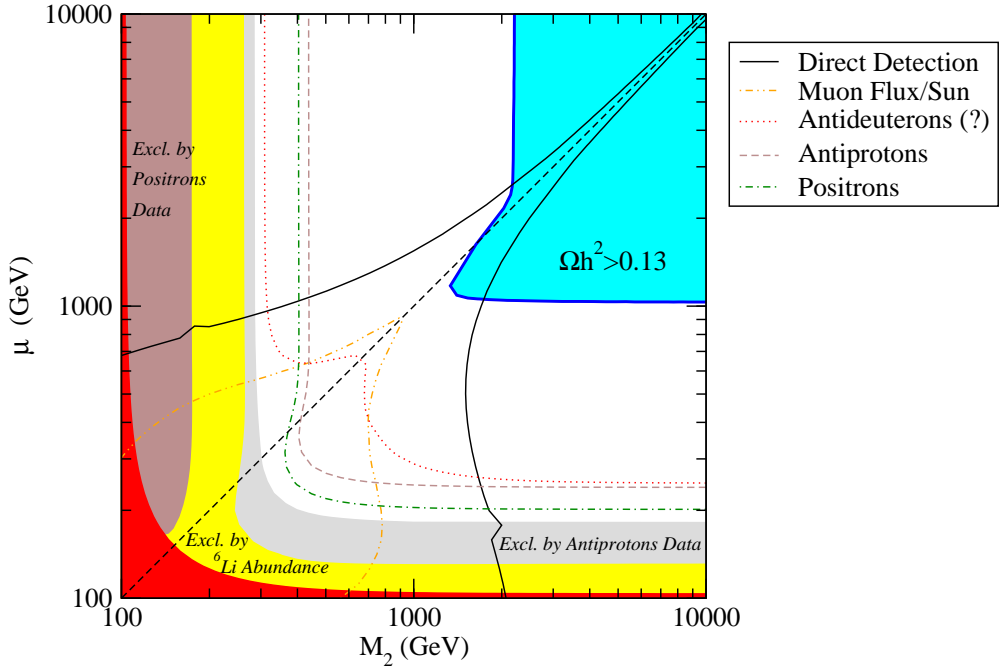}\\
\end{center}
\caption{\small \em A collection of future exclusion limits in
various direct and indirect neutralino detection techniques on the
$(M_2,\mu)$ plane, at $M_1=10$ TeV. A conservative cored dark halo
model (Burkert profile) is assumed. The gray and brown shadings respectively reproduce the parameter space inconsistent with current data on Antiprotons and Positrons. The region shaded in yellow is not consistent with the measured Lithium 6 abundance.}\label{M1summary}
\end{figure*}

Turning to a conservative cored profile, we find again a large antideuterons reach along the Higgsino-Bino large mixing region; in the pure Wino region, instead, the lion's share is given by antiprotons. As a result, the potentially undetectable regions shrink to very narrow corners of massive pure Winos or Higgsinos. We remark, comparing Fig.~\ref{summary1} and \ref{summary2}, a noticeable complementarity among neutralino dark matter searches: within this scenario, in fact, the Higgsino like part of the parameter space would be thoroughly searched for by direct detection experiments, while the pure Wino region, already constrained by current data on antimatter fluxes, will be largely accessible to next generation space based antimatter search facilities. 

The bottom line is therefore that, except for a thin slice at the edge of the Wino-Bino transition line, {\em in the case study we considered here, the whole parameter space will be accessible to future dark matter detection experiments}; this result holds quite independently of the halo model under consideration: a remarkable {\em complementarity among direct searches and antimatter searches} ensures in fact most corners of the parameter space to be within future reach. A naive estimate of the LHC reach, instead, indicates that {\em these models mostly lie beyond the reach of future super-colliders}.

\section{Discussion}
\label{sec:discussion}

In order to summarize the neutralino detection prospects at future dark matter search facilities, we now go back to the parameter space slices of Split SUSY introduced in Sec.~\ref{sec:paramspace}. We will show on those planes our results concerning current and future reaches. We recall that relic abundances inside the WMAP range correspond to the blue contours, and that the regions shaded in red are ruled out by the LEP2 searches for charginos \cite{lep2}. We shade in yellow the region which is not consistent with the ${}^6$Li abundance. In order to be as conservative as possible, we resort here to the cored Burkert profile: 95\% C.L. exclusions limits from current positrons and antiprotons data are also indicated in the Figures.

\begin{figure*}[!h]
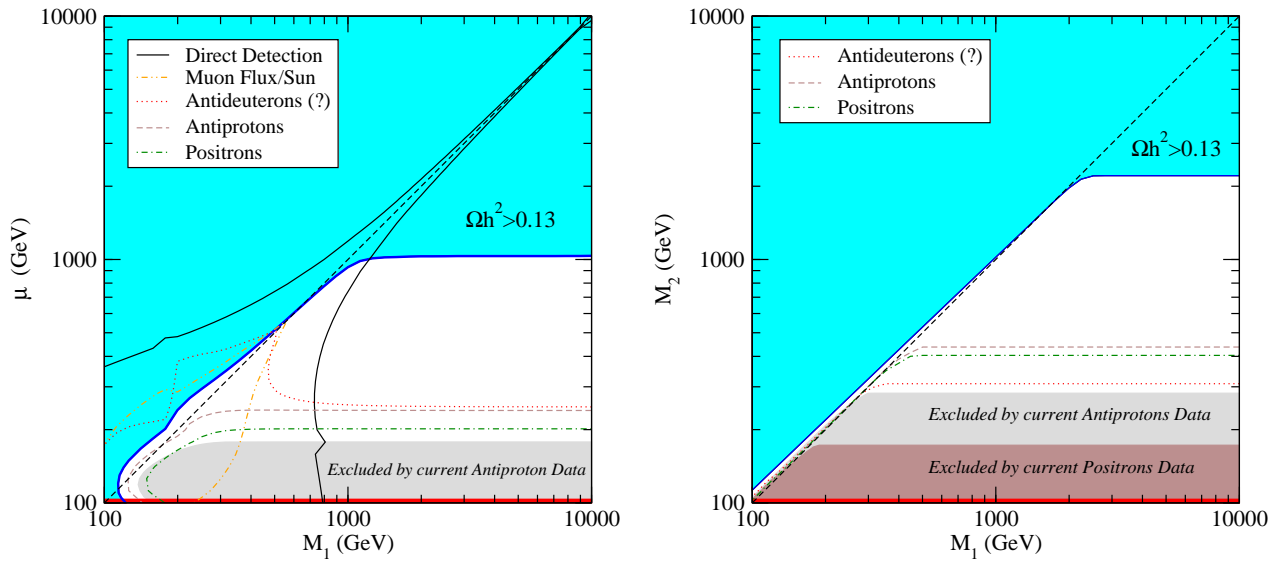

\hspace*{-0.2cm}\includegraphics[scale=0.5]{plots/M2_summary.eps}\quad\includegraphics[scale=0.5]{plots/MU_summary.eps}\\
\caption{\small \em Future exclusion limits on the $(M_1,\mu)$
plane at $M_2=10$ TeV (left), and on the $(M_1,M_2)$ plane at
$\mu=10$ TeV (right). The cored Burkert profile is again assumed
for the dark matter density and velocity
distributions. As in the previous Figure, the gray and brown shadings respectively reproduce the parameter space inconsistent with current data on Antiprotons and Positrons.}\label{M2Musummary}
\end{figure*}

Fig.~\ref{M1summary} details on the $(M_2,\mu)$ plane at large $M_1=10$ TeV. As a first remark, we point out once again the role of Wino-Higgsino mixing, which dramatically enters in direct detection searches: in particular, when the mixing is maximal, masses as large as 10 TeV will be probed at future planned experiments! Antideuterons fluxes, as well as the flux of muons from the Sun, are also sensitive to the degree of mixing between the Wino and Higgsino components in the lightest neutralino: in this setup, IceCube may detect a signal for neutralino masses as heavy as 1 TeV. As regards the case of pure neutralinos, either Wino or Higgsino like, we remark that direct detection techniques are still effective when $M_2,\mu\lesssim$ 1 TeV; on the other hand, if the mixing is tiny, then only antimatter searches are supposed to be effective. We notice, moreover, an interesting and non-trivial complementarity between antiprotons and antideuterons searches: while the first will be more effective in the pure Wino region, the latter are expected to do better in the pure Higgsino case. In general, however, we find that the reach of antimatter searches is limited to around 500~GeV in the Wino like case and to 250~GeV in the Higgsino like one. 

Let us now turn to the $(M_1,\mu)$ and $(M_1,M_2)$ planes, Fig.~\ref{M2Musummary}. In the well-known Bino-Higgsino plane, a large region of sufficiently pure Higgsinos turns out to be excluded by current antiprotons data. As expected, direct detection will probe the region at large mixing, {\em i.e.} when $M_1\simeq \mu$, even at very large masses. As the Higgsino fraction drops (upper part of the Figure), all detection methods are not effective, in this plane. The case of pure Higgsinos reflects what we pointed out above: the region may be best searched for at antimatter detection experiments, up to masses around 250 GeV.
%
%

The $(M_1,\mu)$ plane may be regarded, to some extent, as a low-energy blow-up of the so called {\em hyperbolic branch/focus point region} (HB/FP) of minimal supergravity \cite{hbfp}. In this respect, we can directly compare the results we show in the left panel of Fig.~\ref{M2Musummary} with those recent studies addressing the detection of neutralino dark matter in the HB/FP region \cite{baerdetect,sugrarates}. We remark that, as far as those quantities that weakly depend on assumptions on the halo models are concerned, we mostly agree with those works. In particular, we confirm the importance of the role of neutrino telescopes in the HB/FP region, already pointed out in Baer et al., in Ref.~\cite{baerdetect}: the timeline of the future experimental facilities might give the opportunity to IceCube to provide the first indirect evidences for neutralino dark matter in this region of the mSUGRA parameter space; the would-be signal might later be confirmed by Stage-3 direct detection experiments, whose sensitivity extends well beyond that of the future antartic neutrino telescope.

In the present analysis we extended previous studies of the bino-higgsino mixing region of mSUGRA with a detailed and thorough investigation of indirect neutralino detection through antimatter searches. With a conservative dark halo profile, we find that positrons and antiprotons searches will not be able to probe models, in the HB/FP region, featuring a thermal neutralino relic abundance within the WMAP range. Nevertheless, we point out, as a novel and somewhat exciting result, that antideuterons searches on a GAPS-like experiment could be extremely comptetitive in the HB/FP region, with a reach comparable to that of neutrino telescopes.

The $(M_1,M_2)$ plane, which reproduces a conceivable scenario in which the Gaugino masses are light and the $\mu$ parameter is large, is essentially split in a pure Wino and pure Bino region, respectively below and above the diagonal, dashed line. Interestingly enough, in this plane direct detection and muon fluxes feature extremely low rates, well below future projected sensitivities. The Bino like region, in which most of the points producing a thermal relic abundance in the WMAP range lie, is completely off-limits for any detection technique. The Wino like region, which is already constrained by current data up to $\tilde\chi$ masses as large as 300 GeV, is going to be partly covered by antimatter searches, particularly in the antiprotons channel.

\begin{figure*}[!t]
\begin{center}
\includegraphics[scale=0.59]{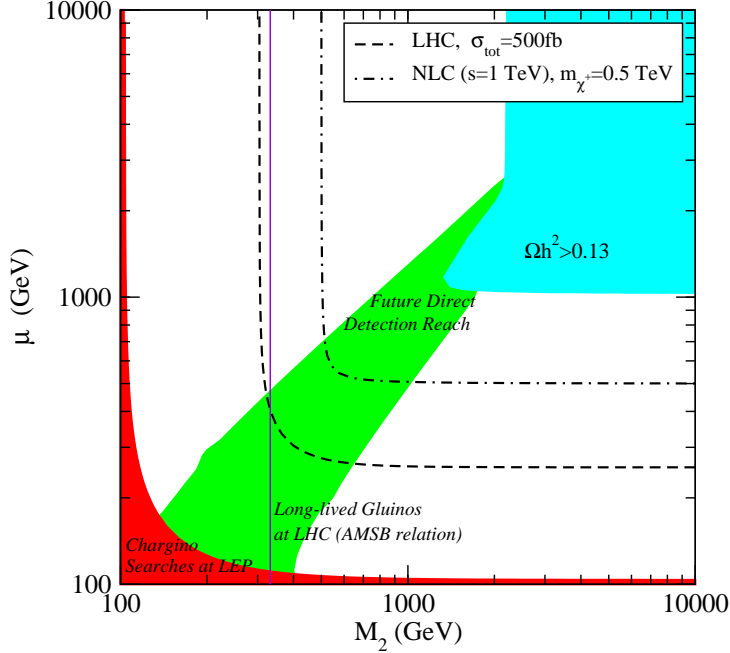}\\
\end{center}
\caption{\small \em The isolevel curve corresponding to the 500 ${\rm fb}^{-1}$ total neutralino-chargino production cross section at the LHC and the (kinematic) reach of a NLC, compared with direct dark matter searches on the
$(M_2,\mu)$ plane, at $M_1=10$ TeV. We {\em rescale} here the local abundance of neutralinos according to their thermal relic density (see Eq.~\ref{eq:rescale}).}\label{summaryrescale}
\end{figure*}

So far we have assumed that, at each point in the parameter space, the lightest
neutralino has a relic density matching the CDM component, regardless of 
its thermal relic density in the standard cosmological framework. We also mentioned 
that relaxing the overproduction bound is possible but contrived, while there is a
plethora of possibilities to enhance a low thermal relic abundance into the 
CDM range. In a more conservative approach one can assume that only a fraction of the CDM is accounted for by supersymmetric models whose thermal relic abundance lies below the WMAP preferred range ($\Omega_\chi<\Omega^{\rm WMAP}_{\rm max}$). Under this assumption, the simplest possible way to compute the detection rates is to rescale the neutralino dark matter density distribution according to the prescription
\begin{equation}\label{eq:rescale}
\rho_\chi(r)\equiv\rho_{\rm CDM}(r)\cdot {\rm min}\left(1,\frac{\Omega_\chi}{\Omega^{\rm WMAP}_{\rm min}}\right).
\end{equation}
Direct detection and capture rates in the Sun linearly scale with the local neutralino density $\rho_\chi(r_0)$; on the other hand, gamma-ray and antimatter rates involve the squared of the neutralino density distributions. We find that the only technique which could give a signal, in this approach, and in the parameter space we analyze here, is the spin-independent direct detection. In Fig.~\ref{summaryrescale} we present the {\em rescaled} direct detection reach in the $(M_2,\mu)$ plane, at $M_1=10$ TeV.
Accelerator searches would naturally not be involved in the above outlined rescaling:
we show in the figure the iso-500 fb total production cross section at the LHC,
and the kinematic reach of a future next linear collider (NLC) with a center-of-mass 
energy of 1 TeV ({\em i.e.} the iso-$m_{\widetilde\chi^+}=500$ GeV line)\footnote{Notice that for nearly degenerate lightest chargino and neutralino the actual reach of a NLC will be slightly lower than the kinematical reach.}.

For completeness, we also mention the detection prospects connected to another
well-known feature of the Split SUSY scenario, i.e. the fact that Gluinos have very
long lifetimes. The phenomenological implications of a quasi-stable Gluino have been recently studied in ref.~\cite{Kilian:2004uj,Hewett:2004nw,Cheung:2004ad}; they mildly depend on the details of the spectrum, one of the crucial quantities being the Gluino mass. 
The maximal LHC reach is around $M^{0}_{\tilde g}\simeq 2$.3 TeV \cite{Kilian:2004uj}.
In the general setup we have considered, the Gluino mass is a free parameter. In order to give a flavor of what could be the LHC reach in the long-lived Gluino detection channel, one has to resort to a particular framework to relate the Gaugino masses. For instance, in the $(M_2,\mu)$ plane it is worthwhile inspecting the consequences of assuming an anomaly-mediated-inspired relation, where $M_2/M_3\approx 6\div7$, the spread being given by RG effects on the $g_2$ and $g_3$ coupling which depend on the value of the gravitino mass parameter $m_{3/2}$ \cite{amsb}. In Fig.~\ref{summaryrescale}  we plot, with a solid violet line, the putative reach of the LHC in the long-lived Gluino detection channel assuming an anomaly mediated relation between $M_2$ and $M_3$, which corresponds to $M_2\simeq 330$ GeV. The latter value is obtained solving for the equation:
\begin{equation}
M_2=\left[\left(\frac{M_3}{M_2}\right)(M_2)\right]^{\rm AMSB}\cdot M^{0}_{\tilde g}, \quad M^{0}_{\tilde g}=2.3\ {\rm TeV}.
\end{equation}

\noindent In summary, Fig.~\ref{summaryrescale} shows that in case the rescaling prescription
is applied, models which could be discovered with the direct detection technique have non-negligible Wino-Higgsino mixing; since the relic abundance of Winos is smaller than that of Higgsinos, at a given mass, we also notice that in this scheme the Higgsino like region is somewhat favored. The pure Wino and pure Higgsino regions could only be explored at accelerators: the complementarity between accelerator searches and the quest for dark matter therefore emerges even in the present setup of pure electroweak production of supersymmetric particles at colliders.

As a concluding remark, we want to stress that the results we presented here hold for more general supersymmetric models, where the scalar sector need not necessarily be largely ``{\em split}''.
We can conservatively state that postulating a scalar sector lying even only one order of magnitude above the LSP mass scale would leave most of our results essentially unaffected. Even lighter scalars may, however, give rise to some possible caveats, particularly for the neutralino relic density, for instance as far as sfermion coannihilations, resonant annihilations with the heavy Higgses, and $t$- and $u$-channels sfermions exchange diagrams are concerned.

\section{Conclusions}
\label{sec:conclusion}

We studied in full generality the neutralino dark matter phenomenology of models in which the scalar sector is heavy, {\em i.e.} in Split Supersymmetry scenarios. The relevant parameters are the entries of the neutralino mass matrix, namely the soft breaking Gaugino mass parameters $M_1$ and $M_2$ and the Higgs mixing mass term $\mu$. Requiring that the thermal relic abundance of neutralinos does not exceed the observed amount of cold dark matter defines an hypersurface in the three-dimensional space of parameters: the cosmological bound is indeed the only issue forcing one of 
the sectors of the theory to be light, {\em i.e.} at a scale lower than at least 
2.2~TeV; the bound can be violated only invoking mechanisms for entropy production
at low energy, a rather contrived setup. Models defined by parameters lying  below the hypersurface are viable, and suitable mechanisms of relic density enhancement may drive low relic density models to produce the required amount of cold dark matter.

We studied in detail the physics of Wino-Higgsino mixing, and pointed out that future experiments may be able to probe extremely large neutralino masses when the mixing is maximal: 10 TeV and 1 TeV neutralinos may give detectable signals respectively in next generation spin-independent searches and at IceCube. Interestingly enough, we find that even though only one Feynman diagram contributes to neutralino-proton scattering, cancellations among the various neutralino interaction-eigenstates components may conspire and suppress the relevant couplings.

Resorting to two extreme cases for the dark matter distribution in our Galaxy, we showed that large parameter space portions are already ruled out by currently available data on antiprotons and positrons fluxes, as well as by the primordial ${^6}$Li abundance induced by neutralino residual annihilations. Depending on the structure of the dark halo in the GC, measurements of the gamma-ray in the Galactic center direction from the EGRET and the H.E.S.S. telescopes have also been shown to set strong constraints
in the parameter space. Future prospects for antimatter searches look promising, particularly in the pure Wino and pure Higgsino cases; a remarkable complementarity between antideuterons and antiprotons detection has also been outlined.

We worked out an explicit analysis of a case study, in which we fixed the Bino mass term $M_1=$0.5 TeV. We showed that while most of the resulting parameter space slice will not be within the reach of the LHC, an interplay among direct detection and antimatter searches will allow future dark matter detection experiments to {\em thoroughly explore} the Split Supersymmetry parameter space.

Restricting to a standard cosmological scenario and taking into account for thermal components only, a rescaling procedure should be implemented for subdominant supersymmetric 
dark matter candidates, suppressing detection rates and leaving a chance only for 
direct detection in the large Wino-Higgsino mixing region. In this scenario, the complementarity among collider searches and direct detection experiments explicitly emerges, although here superpartners can be produced only through electroweak processes.

As a last remark, we point out that a lighter scalar sector would in general yield, modulo cancellations, larger dark matter detection signals: in this respect our results may be regarded as {\em lower bounds} to neutralino searches in more general supersymmetric setups.

\section*{Acknowledgments}  
The work of A.M.\ and P.U.\ was supported by the Italian INFN under the
project ``Fisica Astroparticellare'' and the MIUR PRIN ``Fisica Astroparticellare''.
The work of S.P. was supported in part by  the U.S. Department of Energy under contract number DE-FG02-97ER41022. S.P. would like to thank H.~Baer, K.~Matchev and A.~Birkedal for useful discussions and remarks.\\

\vspace{0.cm}

\noindent {\bf Note added}. While this manuscript was being completed, ref.~\cite{Arvanitaki:2004df} appeared, where some indirect dark matter detection signals are discussed in the framework of Split Supersymmetry.

\end{document}